%% file: main_final.tex
\documentclass[aps,prx,twocolumn,superscriptaddress,english,floatfix,longbibliography]{revtex4-2}
\usepackage{graphicx}
\usepackage{float}
\usepackage{physics}
\usepackage{cancel}
\usepackage{overpic}
\usepackage{enumerate}
\usepackage{dsfont}
\usepackage{xr-hyper}
\usepackage[colorlinks,citecolor=blue,linkcolor=blue,urlcolor=blue,filecolor=black]{hyperref}
\usepackage{textcomp}
\usepackage{amsmath}
\usepackage{amssymb}
\usepackage{soul}
\usepackage[normalem]{ulem}
\usepackage{lipsum}
\usepackage{comment}
\usepackage{mathrsfs}

\usepackage[english]{babel}
\addto{\captionsenglish}{}

\usepackage{bm}
\usepackage{orcidlink}

\graphicspath{{./}{./figures/}}

\renewcommand{\hl}[1]{{#1}}

\input{./shortcuts}

\begin{document}

	\title{Monitored long-range interacting systems: spin-wave theory for quantum trajectories}
	
	\author{Zejian Li~\orcidlink{0000-0002-5652-7034}}
    \email{li.zejian@ictp.it}
	\affiliation{The Abdus Salam International Center for Theoretical Physics, Strada Costiera 11, 34151 Trieste, Italy}
	\author{Anna Delmonte~\orcidlink{0009-0008-9371-6855}}
	\affiliation{SISSA, Via Bonomea 265, I-34136 Trieste, Italy}
	\author{Xhek Turkeshi~\orcidlink{0000-0003-1093-3771}}
	\affiliation{Institut f\"ur Theoretische Physik, Universit\"at zu K\"oln, Z\"ulpicher Strasse 77, 50937 K\"oln, Germany}
	
	\author{Rosario Fazio~\orcidlink{0000-0002-7793-179X}}
	\affiliation{The Abdus Salam International Center for Theoretical Physics, Strada Costiera 11, 34151 Trieste, Italy}
	\affiliation{Dipartimento di Fisica ``E. Pancini", Universit\`a di Napoli ``Federico II'', Monte S. Angelo, I-80126 Napoli, Italy}
	
	\begin{abstract}
		Measurement-induced phases exhibit unconventional dynamics as emergent collective phenomena, yet their behavior in tailored interacting systems -- crucial for quantum technologies -- remains less understood. We develop a systematic toolbox to analyze monitored dynamics in long-range interacting systems, relevant to platforms like trapped ions and Rydberg atoms.
Our method extends spin-wave theory to general dynamical generators at the quantum trajectory level, enabling access to a broader class of states than approaches based on density matrices.
This allows efficient simulation of large-scale interacting spins and captures nonlinear dynamical features such as entanglement and trajectory correlations. 
We showcase the versatility of our framework by exploring entanglement phase transitions in a monitored spin system with power-law interactions in one and two dimensions, where the entanglement scaling changes from logarithm to volume law as the interaction range shortens, and by dwelling on how our method mitigates experimental post-selection challenges in detecting monitored quantum phases. 
		
	\end{abstract}
	\maketitle

	\section{Introduction}
	
	Long-range interacting quantum many-body systems have recently been the focus of intense theoretical and experimental activity \cite{RMPDefenu1,RDefenu2}. 
	Sufficiently non-local interactions between the elementary constituents of a many-body system lead to exotic non-equilibrium phenomena, with compelling signatures in the properties of quantum correlations and entanglement, relaxation  dynamics, quantum information scrambling, and ergodicity-breaking properties.
	Along with their importance in statistical mechanics, long-range quantum systems are central to the rising field of quantum technologies and simulators. Experimental platforms such as trapped ions~\cite{Monroe_2021}, Bose-Einstein condensates in cavities~\cite{Ritsch}, dipolar~\cite{Lahaye_2009}, polar~\cite{Bohn_2017}, and Rydberg atoms~\cite{Weimer_2010}, or driven ultra-cold atomic gases~\cite{ferioliNonequilibriumSuperradiantPhase2023} showcase frameworks where the system interactions scale as a power law of the distance.

	The dynamics of long-range systems has been investigated in the two limits of unitary and dissipative (Lindbladian) evolutions. In this work, we would like to take a step forward by studying the in-between framework of monitored dynamics.
	Here, the system evolution is interspersed with quantum measurements, whose outcomes are stochastic. 
	As a result, the system is described by a quantum trajectory conditional to the measurement registry. Averaging the state over the trajectory ensemble recasts a dissipative dynamics, therefore presenting a convenient framework for the study of physical observables in Lindblad dynamics~\cite{Plenio_1998,Wiseman,Carmichael}. 
	It was more recently realized, in the study of many-body systems, that the quantum trajectory ensemble contains richer information than the mean state, showcased by the several collective phenomena encoded in beyond-average statistical features. The cornerstone examples are monitored quantum phases and measurement-induced phase transitions~\cite{Skinner_prx,Li_Fisher,Cao}, characterized by non-linear functions of the trajectories, such as entanglement probes or trajectory correlation functions.
	Intensive work gathered insights on the monitored phases in local random quantum circuits~\cite{Li_Fisher2,zabalo202criticalpropertiesof,zabalo2022operatorscalingdimensions,sierant2022universal,sierant2022measurementinducedphase,PhysRevX.12.011045} and non-interacting systems~\cite{Fava,Turkeshi,poboiko2023theoryoffree}, while 
	few works have addressed monitored dynamics in long-range interacting systems. Refs.~\cite{Block,sharmaMeasurementinducedCriticalityExtended2022} studied Clifford circuits with two-qubit gates entangling distant sites with a probability of decaying as a power-law of their distance.
	The Floquet dynamics of interacting spin systems with measurements and feedback was considered in a model with entanglement and dissipative phase transitions~\cite{Sierant2022dissipativefloquet}, while monitored long-range free fermions have been studied in Refs.~\cite{Muller,Minato}. All these results demonstrate that long-range couplings strongly affect the monitored phases and their transitions: they are relevant in the renormalization group sense.

	Experimentally detecting monitored phases, on the other hand, is challenging, and so far limited to few  pioneering works on trapped-ions~\cite{noel2021measurementinducedquantum} and superconducting platforms~\cite{
		KohIBM,Google1}.
	This limitation has a fundamental origin known as the \emph{post-selection problem}. 
	Observing beyond-average moments of the quantum trajectories requires reproducing the same sequence of measurement outcomes multiple times to collect enough statistics and obtain trajectory-average values.
	This task is formidable, as the probability of reproducing the same trajectory is exponentially small in system size and time scale.

	The post-selection barrier is in general ineludible for the study of generic many-body monitored systems, and the quest for methods that can mitigate it is an active research line.
	For instance, the presence of feedback dynamics may be designed to imprint the measurement-induced transition into the density matrix~\cite{PhysRevLett.131.060403,buchhold2022revealing}, albeit in general leads to separate types of phase transitions~\cite{ravindranath2022entanglementsteeringin,odea2022entanglementandabsorbing,piroli2022trivialityofquantum,sierant2023controlling,allocca2024statistical,lemaire2023separate,Sierant_2023}.
	A complementary path is available for systems that can be classically simulated. 
	When the post-processing is efficient and faithful, it allows to elude the post-selection~\cite{gullans2020scalableprobesof,Li_2023,li2021robustdecodingin,garratt2023probing,garrattMeasurementsConspireNonlocally2023,kamakari2024experimental,yanay2024detecting}. 
	There are special cases in which the system itself is designed to be immune to post-selection. In~\cite{Passarelli} some of us showed that there is a class of infinite-range spin systems where monitored many-body dynamics can be efficiently realized with a post-selection overhead scaling, at most, as a power of the system size. 
	It was argued that this fortunate case was not specific to that model, 
	but it applies to a broad class of monitored systems with an underlying semi-classical dynamics.

	A central target of this work is supporting the above claims for 
	sufficiently long-range interacting spin systems. 
	This goal is achieved by
	\begin{itemize}
		\item developing 
		a systematic semi-classical expansion for quantum trajectories that is particularly suited for long-range systems,
		\item showing  
		that, in this regime, the post-selection barrier is avoidable, paving the way for future experiments. Trapped ions, among others, belong to the systems for which our method and results apply.
	\end{itemize}
	
	{As long-range systems are many-body and interacting by construction, their theoretical study also poses a formidable challenge by itself. One may resort to approximative numerical methods, such as those based on matrix-product states~\cite{Doggen,Daley_2014,jaschkeOnedimensionalManybodyEntangled2018a} and other types of tensor networks~\cite{sulzNumericalSimulationLongrange2024a}, representing the state-of-the-art for one-dimensional systems and in low entropy 
 scenarios. These methods, while being ``generic'', struggle in capturing correlations built up in nonequilibrium dynamics (for example in a phase transition) and generally suffer from poor complexity scaling in two or higher spatial dimensions. The method we propose, on the other hand, exploits the simplifications granted by long-range interactions: they  typically lead to underlying semi-classical dynamics, allowing 
	quantum excitations and fluctuations to be treated perturbatively on top of a classical model.}
		These observations lead to a class of approximation methods known as the \textit{spin-wave} theory. 
	Historically, the spin-wave theory was first introduced by Bloch~\cite{blochZurTheorieAustauschproblems1932} in 1932 for ferromagnetic spin systems in equilibrium. An equivalent formulation of the theory was later given by Holstein and Primakoff~\cite{holsteinFieldDependenceIntrinsic1940}. The
	spin-wave theory has been highly successful in describing equilibrium ground states in a wide range of magnetic materials, showing good accordance with experimental data~\cite{vankranendonkSpinWaves1958}. More recently, it has been generalized to the time-dependent regime~\cite{ruckriegelTimedependentSpinwaveTheory2012a,leroseImpactNonequilibriumFluctuations2019}, capturing non-equilibrium dynamics in closed quantum systems undergoing unitary time evolution governed by a Hamiltonian and in driven-dissipative systems~\cite{seetharamDynamicalScalingCorrelations2022} described by a Lindblad master equation.
	
	This work enlarges spin-wave theory to encompass generalized measurements, therefore providing a natural framework for the study of monitored Hamiltonian and Lindbladian long-range systems. {We demonstrate the method on both one- and two- dimensional spin lattices and on a spin-boson model, exploring different monitored phases where the entanglement scaling ranges from logarithmic to volume law.} Moreover, as our method is designed on the level of single quantum trajectories, it also provides a much more accurate representation of the Lindblad dynamics, {allowing to capture non-Gaussian corrections that are by default lost in semiclassical} approaches {approximating directly} on the level of the density matrix~\cite{seetharamDynamicalScalingCorrelations2022}, offering a broad spectrum of applications.

	The paper is organized as follows. In Section \ref{sec:model} we set the stage by defining the class of models we are going to examine and the associated Lindblad equation. Section~\ref{sec:swqt} is the core of our work as it contains the essence of the stochastic spin-wave approach. In this Section, we explain how it is constructed in the case of a monitored dynamics described by a quantum state diffusion.
	The method is then applied in Section~\ref{sec:ept-long-range} where we benchmark it with exactly solvable cases and we apply it to the study of dissipative phase transitions and entanglement phase transitions in long-range spin systems. 
	In Section~\ref{sec:experimental} we discuss how the post-selection problem can be avoided in long-range spin systems by exploiting quantum-classical cross-correlated observables enabled by our spin-wave framework. Finally, we draw our conclusions in Section~\ref{sec:conclusion}. Some technical details are included in the Methods section.
	
	\begin{figure*}[t]
		\centering
		\includegraphics[width=\linewidth]{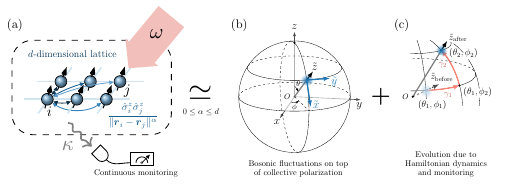}
		\caption{Sketch of a long-range spin model [panel (a)] and the representation of its monitored dynamics in the spin-wave approximations [panels  (b) and (c)]. (a) The spins on a {$d$-dimensional lattice} are collectively driven (at amplitude $\omega$) and interact via $z-z$ coupling, whose strength decays as a power law of the distance (with exponent $\alpha$). The system is subjected to a collective dissipation (at rate $\kappa$) and continuously monitored. In the {infinite-range or} long-range regime ({$0
			\leq\alpha\leq d$}), the state of the system can be represented in the spin-wave approximations [panel (b)], where spin fluctuations around the collective polarization are bosonized. Here, $Oxyz$ depicts the lab frame and $O\xtil\ytil\ztil$ is the instantaneous frame where $\ztil$ aligns with the collective spin polarization. Panel (c) represents the evolution of the state under both the Hamiltonian dynamics and the continuous monitoring. In the proposed method, the angles $(\theta,\phi)$ of the instantaneous frame $O\xtil\ytil\ztil$ are updated to match the collective spin direction after its evolution in each infinitesimal time step. The dashed (solid) straight arrow represents the $\ztil$ axis before (after) the re-alignment step and the curved arrows represent the paths $\gamma_1$ and $\gamma_2$ as defined in Eq.~\eqref{eq:integral-paths} for evaluating the integral of Eq.~\eqref{eq:beta-align-pde}. }
		\label{fig:power-law-spins}
	\end{figure*}
	
	\section{Results}
	
	\subsection{Monitored long-range spin systems}
	\label{sec:model}
	
	We consider a system with $N$ spin-$s$ degrees of freedom on a lattice whose sites are labeled by indices $i,j$, that is subjected to continuous monitoring. For concreteness, we consider the case of weak monitoring where the dynamics is governed by the following stochastic master equation (in units where $\hbar=1$),
	
	\bea\label{eq:het-eq-rho}
	\d{\hat\rho} &= \dt\Lcal(\rhohat) \\
	&+ \sum_i\[ \d \mbox{w}^*_i\left(\Lhat_i-\langle \Lhat_i \rangle\right)\rhohat  + \d \mbox{w}_i\rhohat\left(\Lhat^{\dagger}_i-\langle \Lhat^{\dagger}_i \rangle\right)\]\,,
	\eea
	where $\rhohat$ is the density matrix, $\mathcal{L}$ is the Liouvillian superoperator 
	\bea
	\Lcal(\rhohat)  \equiv -\rmi[ \Hhat, \rhohat] + \dcal (\rhohat) \,,\label{eq:liouville}
	\eea 
	$\hat{L}_i$ {is the} Lindblad operator acting on the $i$-th spin, and $\langle\Lhat_i\rangle\equiv\tr[\rhohat\Lhat_i]$ denotes the (single-trajectory) expectation value. In Eq.~\eqref{eq:liouville}, 
	$\Hhat$ is the Hamiltonian generating the coherent dynamics, whose form will be specified later, and $\dcal$ denotes the superoperator governing the dissipation, that acts on the density matrix as follows,
	\bea
	\label{eq:dissipator-fmn}
	\dcal(\rhohat) &\equiv \sum_{i,j}f_{ij}\left( \Lhat_i\rhohat\Lhat_j^{\dagger} - \dfrac{1}{2}\left\{\Lhat^{\dagger}_j\Lhat_i,\rhohat\right\}\right)\, .
	\eea
	The positive semidefinite matrix $f_{ij}$ takes into account spatial correlations among the Lindblad operators $\Lhat_i$. 
	In the stochastic master equation~\eqref{eq:het-eq-rho}, $\d \mbox{w}_i$ is a complex Wiener process satisfying the relations
	\bea\label{eq:dw}
	& \overline{\d \mbox{w}_i} = 0\,, &  \\
	& \d \mbox{w}_i^* \d \mbox{w}_j  = f_{ij}\d t\,,~ \d \mbox{w}_i \d \mbox{w}_j =0\,,
	\eea
	where the $\overline{\phantom{|}\bullet\phantom{|}}$ notation denotes the ensemble average.
	Note that in the general case where $f_{ij}$ is nondiagonal, the noises are also spatially correlated. 
	
	We consider the dynamics of a pure state $\rhohat = |\psi\rangle\langle \psi|$. Given the purity preservation of the dynamics~\eqref{eq:het-eq-rho} (cf. Supplementary Note~\ref{app:nondiagonal-unravel}), the von Neumann entropy
	\bea
	S_E \equiv -\tr \[\rhohat_{\frac{N}{2}}\log \rhohat_{\frac{N}{2}}\]\,
	\eea
	is a good quantifier of entanglement~\cite{vidalEntanglementQuantumCritical2003}. Here, $\rhohat_{\frac{N}{2}}\equiv\tr_{\{1,\ldots,\lfloor N/2\rfloor\}}[|\psi\rangle\langle \psi|]$ denotes the half-system reduced density matrix (in a bipartition where one subsystem contains spins indexed from $1$ to $\lfloor N/2\rfloor$).

	Using Eqs.~\eqref{eq:het-eq-rho} and~\eqref{eq:dw}, one verifies that the trajectory-average state (which we also denote by $\rhohat$ when there is no confusion with the single-trajectory state) evolves deterministically according to the Lindblad master equation
	\bea\label{eq:swqt-master-eq}
	\dfrac{\d}{\d t}\rhohat = \Lcal(\rhohat)\,.
	\eea
	As shown in the Supplementary Note~\ref{app:nondiagonal-unravel}, Eq.~\eqref{eq:het-eq-rho} is an unraveling of the master equation~\eqref{eq:swqt-master-eq} and describes the quantum state diffusion process subjected to a heterodyne-detection monitoring scheme~\cite{gisin1992thequantumstate,Wiseman}.

	\subsubsection{Power-law spin model}\label{sec:power-law-spin-model}
	
	While the formulation of our spin-wave method will not depend on the particular form of the model, we generally require $\Hhat$ and $\dcal$ to be long-range in order for the method to yield an accurate approximation to the exact solution. For concreteness, we study a prototypical driven-dissipative spin model with spatially extended interaction whose strength decays as a power-law according to the distance, as sketched in Fig.~\ref{fig:power-law-spins} (a). The model is defined on a {$d$}-dimensional periodic {lattice of $N=L^d$ spins at positions $\vec{r}_i\equiv (r_i^{(1)},\ldots,r_i^{(d)})$ where $r_i^{(p)} = 1,\ldots, L$}.  {The}  Hamiltonian {is}
	\bea\label{eq:ham-btc-zz}
	\Hhat = \omega\Shat^x+\dfrac{2sJ}{\ncal}\sum_{i, j}\dfrac{\sigmaz_i\sigmaz_j}{{\lVert\vec{r}_i-\vec{r}_j\rVert}^\alpha}\,,
	\eea
	where $\omega$ is the amplitude of a collective drive and $J$ is the interaction strength. The factor $2s$ ensures that the mean-field theory (see Supplementary Note~\ref{app:mf-power-law-spin}) is $s-$independent, which can be ignored for spin-half ($s=1/2$) systems.
	Here, we denote total spin operators with the capital $\Shat^\mu$, $\mu\in\{x,y,z\}$ without the site index subscript  $\Shat^\mu\equiv\sum_i\shat_i^\mu\,$,
	while the total spin number is denoted by     $S \equiv Ns$.
	The lower-case notation $\shat_i^\mu$ refers to the spin operator of the $i-$th site, which satisfies the standard $\mathfrak{su}(2)$ algebra
	$[\shat_i^\mu,\shat_j^\nu]=\rmi\delta_{ij}\epsilon^{\mu\nu\gamma}\shat^\gamma$.
	We denote the normalized spin operators with
	$\sigmahat_i^\mu\equiv \shat^\mu_i/s\,$,
	which reduce to the standard Pauli matrices in the case of spin-half. The distance on the periodic {lattice} is 
	{$\lVert \vec{r}_i - \vec{r}_j\rVert\equiv  \sqrt{\sum_{p=1}^d\min(|r_i^{(p)}-r_j^{(p)}|,L-|r_i^{(p)}-r_j^{(p)}|)^2}$}, 
	and in the case of $i=j$ we adopt the convention of ${\lVert \vec{r}_i - \vec{r}_i\rVert}=\infty$ such that there is no on-site self-interaction for finite $\alpha$.
	The Kac normalization~\cite{kacVanWaalsTheory1963}  
	$\ncal\equiv\frac{1}{N}\sum_{i,j}{\lVert \vec{r}_i - \vec{r}_j\rVert}^{-\alpha}\,$,
	ensures a well-defined thermodynamic limit for the interaction Hamiltonian.
	The power $\alpha$ determines the range of the interaction. In particular, the case of $\alpha=0$ describes an infinite-range model with permutation invariance, and the opposite limit $\alpha\to\infty$ corresponds to an Ising model with nearest-neighbor interaction. The long-range regime corresponds to the case where {$0<\alpha\leq d$}~{\cite{campaStatisticalMechanicsDynamics2009}}. 
	
	In addition to the unitary dynamics governed by the Hamiltonian, we subject the spin chain to a collective (infinite-range) decay, a case of experimental relevance~\cite{ferioliNonequilibriumSuperradiantPhase2023, ferioliLaserDrivenSuperradiantEnsembles2021}, resulting in the following Lindblad master equation,
	\bea\label{eq:zz-master-eq}
	\dfrac{\d\rhohat}{\d t} = -\rmi [\Hhat, \rhohat] + \dfrac{\kappa}{S}\(\Shat^-\rhohat\Shat^+-\dfrac{1}{2}\left\{\Shat^+\Shat^-,\rhohat\right\}\)\,,
	\eea
	with $S$ serving the role of Kac normalization for the dissipator, $\kappa$ is the dissipation strength, and $\Shat^{\pm}\equiv \Shat^x\pm\rmi\Shat^y$
	are the collective spin raising and lowering operators. Note that this collective dissipator corresponds to the choice of $f_{ij}=\kappa/S=\mathrm{const}$ and $\Lhat_i=\shat_i^-=\shat_i^x-\rmi\shat_i^y$ in terms of our generic notation in Eq.~\eqref{eq:dissipator-fmn}, which also fixes the stochastic unraveling via Eq.~\eqref{eq:het-eq-rho}.
	
	The quantum state of the spin system lives in a $2^N$-dimensional Hilbert space, making the exact solution intractable in practice for thermodynamically large $N$. In the following section, we elaborate the theory of spin-wave quantum trajectories, allowing us to overcome the above limitations and tackle the problem in the thermodynamic limit. 
	
	\subsection{Spin-wave theory along quantum trajectories}\label{sec:swqt}
	
	In this section, we present the framework of spin-wave quantum trajectories (SWQT), that is the main result of our work.  This formalism serves as a semi-classical method for solving quantum trajectories of generic out-of-equilibrium dissipative/monitored spin systems with sufficiently long-range interactions whose average dynamics can be described by a Lindblad master equation. The resolution of \emph{single quantum trajectories} enables us to probe non-linear correlations, encoded in the entanglement and other non-linear functions of the state~\cite{verstraelenQuantumClassicalCorrelations2023}.

	The central assumption is that the system admits a strong collective spin polarization, on top of which spin-wave excitations are bosonized via a Holstein-Primakoff expansion truncated to the lowest order:
	\bea\label{eq:bosonization-n}
	\shat^{\ztil}_i(\theta,\phi) &= s - \dbbb_i\bbb_i\,,\\
	\shat^{\xtil}_i(\theta,\phi) &\simeq  \sqrt{\dfrac{s}{2}}(\dbbb_i + \bbb_i)\,,\\
	\shat^{\ytil}_i(\theta,\phi) &\simeq  \rmi\sqrt{\dfrac{s}{2}}(\dbbb_i - \bbb_i)\,.
	\eea
	Here, the $\shat^{\alphatil}$'s are spin operators in the rotated frame $O\xtil\ytil\ztil$ as illustrated in Fig.~\ref{fig:power-law-spins} (b). They are related to the lab-frame spin operators via $\shat_i^\alphatil(\theta,\phi) = \uhat(\theta,\phi)\shat_i^\alpha\uhat^\dagger(\theta,\phi)$ with $ \uhat(\theta,\phi) = \rme^{-\rmi \phi \Shat^z}\rme^{-\rmi\theta\Shat^y}$. The $\ztil$ axis of the rotated frame is aligned with the collective (average) polarization of all the spins, which can be met by imposing $\langle\Shat^\xtil\rangle = \langle\Shat^\ytil\rangle =0\,$. The $\bbb_i$'s are bosonic operators, representing spin fluctuations around the collective magnetization, with standard bosonic commutation relations $[\bbb_i,\dbbb_j]=\delta_{ij}$. This transformation effectively approximates the Bloch sphere with the tangent plane at the north pole of the rotated frame~\cite{RDefenu2}, and becomes exact only when $\langle\dbbb_i\bbb_i\rangle=0$, i.e., when the system is in a spin coherent state. Thus, the density of bosonic excitations
	\bea\label{eq:boson-density-sum-n}
	\epsilon \equiv \dfrac{1}{Ns}\sum_i \langle\dbbb_i\bbb_i\rangle\,,
	\eea
	also referred to as \textit{spin-wave density},
	serves as a natural control parameter for the approximation~\cite{seetharamDynamicalScalingCorrelations2022}.

	The bosonic modes are then approximated with a Gaussian ansatz~\cite{verstraelen2018gaussian,verstraelenGaussianTrajectoryApproach2020a} parametrized by first and second moments.
	This allows us to uniquely specify the state of the entire system with the following variational parameters:
	\bea\label{eq:var-params}
	\theta\,,\phi\,, \beta_i\equiv\langle\bbb_i\rangle\,, u_{ij}\equiv \langle \delhat_i\delhat_j\rangle\,, v_{ij}\equiv \langle \ddelhat_i\delhat_j\rangle\,,
	\eea
	where $\delhat_i\equiv\bbb_i-\beta_i$. Therefore, this method requires only $\ocal(N^2)$ parameters to represent the state in the most general case, which is exponentially efficient as compared to the dimension of the full Hilbert space $2^N$, and the complexity can be further reduced in the presence of additional symmetries. 
	
	The stochastic master equation~\eqref{eq:het-eq-rho} then translates into the dynamics of the variational parameters~\eqref{eq:var-params}. Our algorithm is based on the Euler–Maruyama method~\cite{kloeden1992stochastic}, where time is discretized into small steps $\delta t$.
	Initializing the parameters at $t=0$ such that the rotated frame parametrized by $(\theta,\phi)$ has its $\ztil$ axis aligned with the collective spin (which implies $\sum_i\beta_i = 0$), we proceed stroboscopically by repeating the two following steps until the desired time $t$ is reached:
	\begin{enumerate}
		\item Calculate the infinitesimal increments for the Gaussian parameters $\delta\beta_i$, $\delta u_{ij}$ and $\delta v_{ij}$ using Eq.~\eqref{eq:het-eq-rho}, and update these quantities with the increments. Note that the frame angles $(\theta,\phi)$ are kept constant within this step, which implies that the $\ztil$ axis no longer aligns with the collective spin after the update as $\sum_i\delta \beta_i\neq 0$ in general.
		
		\item Update frame angles $(\theta,\phi)$ such that $\sum_i\beta_{i}=0$ in the new rotated frame [see Fig.~\ref{fig:power-law-spins} (c) for a schematic representation of this step]. This can be achieved self-consistently by considering the evolution of the bosonic mode $\bbb_i$ generated by the (passive) rotation of the frame alone. Then update the Gaussian parameters $\beta_{i}$, $u_{ij}$ and $v_{ij}$ accordingly (due to the rotation of the frame). Finally, increase the time by $\delta t$ and start a new iteration.
	\end{enumerate}
	We detail in Sec.~\ref{sec:update-rules} the key ingredients for performing the two steps sketched above. {Note that as this construction does not depend on the spatial structure of the system, it is straightforward to apply the method to arbitrary dimensions and to adapt to different lattice geometries, without comprising its computational efficiency. Moreover, we would like to stress that the framework constructed here is fully generic, as it suffices to systematically apply the bosonization and the Gaussian approximation to obtain the equations of motion. Therefore, it can be straightforwardly extended to other types of systems beyond those we considered. An example of its generalization to spin-boson systems is presented in Supplementary Note~\ref{sec:spin-boson-theory} together with some illustrative numerical results.}

	Finally, let us remark 
	that since our approximations are formulated at the level of single trajectories and not the average dynamics, our variational ansatz allows us to reach a larger class of density matrices~\cite{verstraelenGaussianTrajectoryApproach2020a} compared to previous methods working on the level of the density matrix~\cite{seetharamDynamicalScalingCorrelations2022}. 
	Indeed, the mixture of Gaussian states, such as those describing single trajectories, is a non-Gaussian state in general, hence encodes non-trivial correlations between the spin-waves.
	This fact has a fundamental operational consequence: spin-wave quantum trajectories provide more accurate representations of the purely Lindblad dynamics, making the SWQT framework compelling also for the study of the dissipative dynamics of hermitian operators. 
	
	\subsection{Entanglement phase transition in the power-law spin model}\label{sec:ept-long-range}
	
	In this section, we apply the presented framework of spin-wave quantum trajectories to study the power-law interacting spin model introduced in Sec.~\ref{sec:power-law-spin-model}. We focus on $s=1/2$ and fix the interaction strength at $J=0.1\kappa$ for the rest of the section.

	\subsubsection{Infinite-range case: $\alpha = 0$}\label{sec:infinite-range}
	
	We now test the validity of the approximations by evaluating both linear (e.g., observables) and non-linear (e.g., entanglement entropy) functions of the state. In order to benchmark the method, we first revisit the known case of $\alpha=0$, where the $z-z$ interaction is all-to-all with no spatial resolution. In this regime, the system can be effectively represented as a single spin $S=N/2$, allowing us to benchmark the spin-wave method against the exact simulation of the dynamics in the Dicke basis. 
	
	\begin{figure*}
		\centering
		\includegraphics[width=\linewidth]{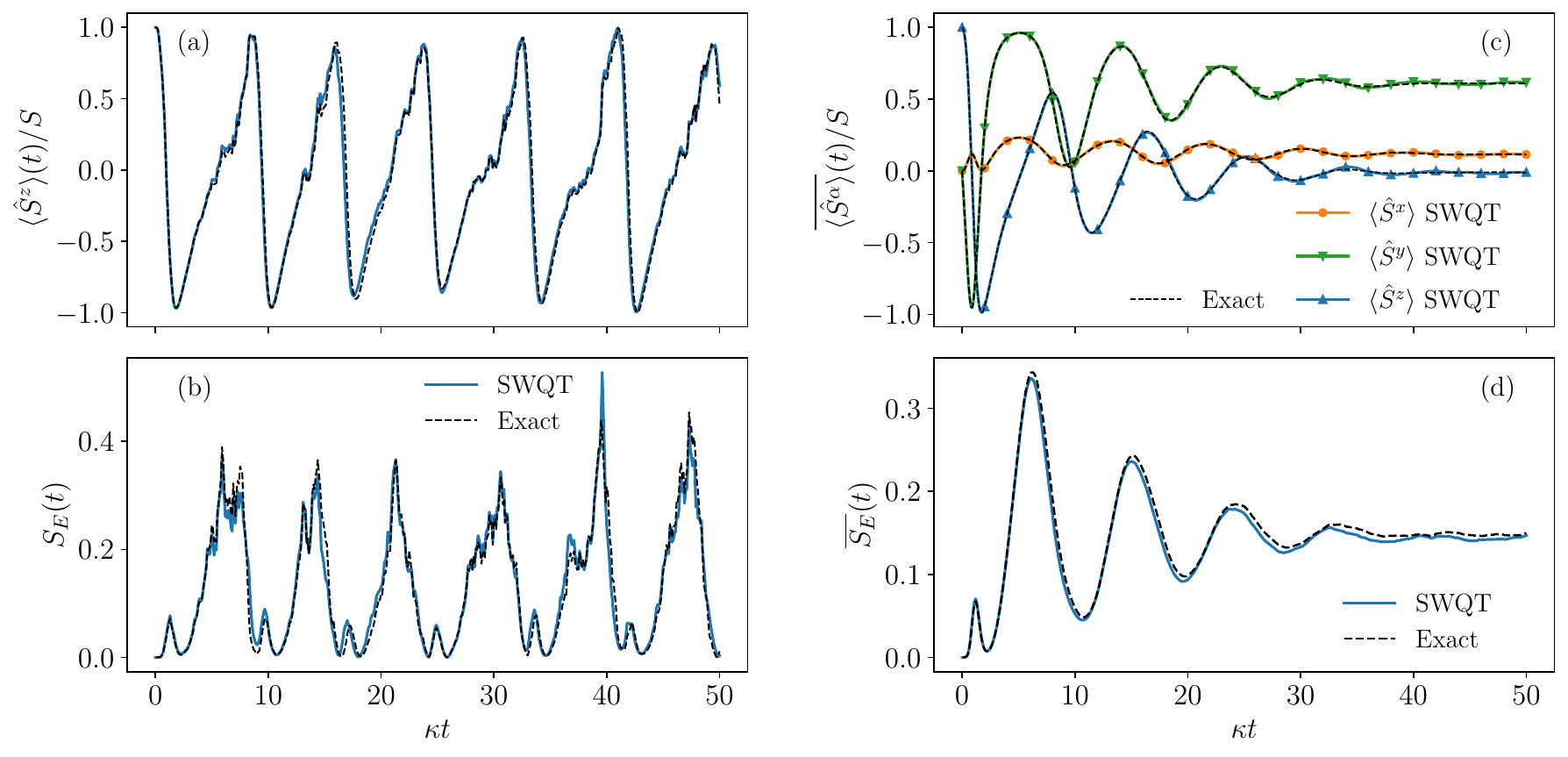}
		\caption{Benchmark of the spin-wave method against the exact solution for the time-evolution of the following quantities: (a) the magnetization $\langle\Shat^z\rangle$ along a single trajectory; (b) the half-system entanglement entropy $S_E$ along the same trajectory as in (a); (c) trajectory-average of the collective spin vector $\overline{\langle\Shat^\alpha\rangle}$; (d) trajectory-average of the half-system entanglement entropy $\overline{S_E}$. In single-trajectory benchmarks [(a) and (b)], the same noise realization is adopted by both the spin-wave quantum trajectory and the exact integration of the stochastic master equation. The trajectory-average benchmarks [(c) and (d)] are performed over 4000 trajectories. Parameters: $\omega=1.25\kappa$, $J=0.1\kappa$,  $S=64$, $\kappa\delta t=10^{-4}$.}
		\label{fig:bench-single-and-avg}
	\end{figure*}

	To put the spin-wave quantum trajectories in fair comparison with the exact ones, we integrate the exact stochastic master equation~\eqref{eq:het-eq-rho} using the Euler-Maruyama method, adopting the same time step and the same noise realization as used in the spin-wave calculation. An example of a single-trajectory benchmark for the spin-wave method is shown in Fig.~\ref{fig:bench-single-and-avg} (a) and (b), where we compare trajectories for both the magnetization $\langle\Shat^z\rangle$ and the half-system entanglement entropy (cf.~Supplementary Note~\ref{app:gaussian-entanglement} and Ref.~\cite{leroseOriginSlowGrowth2020, serafiniSymplecticInvariantsEntropic2003}) $S_E$ against exact ones. The considered system has $S=64$ and $\omega=1.25\kappa$, which, as predicted by the mean-field theory, corresponds to the time-crystal phase in the thermodynamical limit. The initial state is the fully polarized Dicke state (all spins pointing up) for both simulations. Surprisingly, the trajectories obtained with the spin-wave method faithfully reproduce the dynamics of both quantities throughout most of the evolution. This suggests that the proposed method can be used as an approximation to resolve dynamics on the level of single trajectories.

	We also evaluate the performance of the spin-wave method on trajectory-averaged quantities. Fig.~\ref{fig:bench-single-and-avg} (c) shows the time-evolution of the spin expectations $\overline{\langle\Shat^{x,y,z}\rangle}$ obtained by averaging spin-wave quantum trajectories. These are then benchmarked against the exact solution of the Lindblad master equation~\eqref{eq:zz-master-eq}. As expected from the single-trajectory performance, the average dynamics is accurately reproduced. Notably, the decaying oscillations of the magnetization are due to the finite size effect and therefore require corrections beyond the mean-field level to capture. It is interesting to note that the length of the averaged collective spin vector $\overline{\langle\Shat^\alpha\rangle}$ is only a fraction of the maximal value $S$, which implies the that the average density matrix, of highly mixed nature, is far from a state that can be well approximated by the spin-wave approximations.
	As a consequence, a deterministic version of the spin-wave theory with the same approximations applied to the averaged state, e.g., the one proposed in Ref.~\cite{seetharamDynamicalScalingCorrelations2022}, will completely fail to capture the dynamics in this regime, cf. Supplementary Note~\ref{app:swqt-vs-dsw}. In Fig.~\ref{fig:bench-single-and-avg} (d), we compare the trajectory-averaged entanglement entropy with that calculated from exact trajectories, which shows a good agreement again. We also report the average spin-wave density to be $\overline{\epsilon}\lesssim 10^{-2}$ throughout the time evolution (see Supplementary Note~\ref{app:benchmar-steady-state} for a detailed discussion and additional benchmarks).

	\begin{figure}
		\centering
		\includegraphics[width=\linewidth]{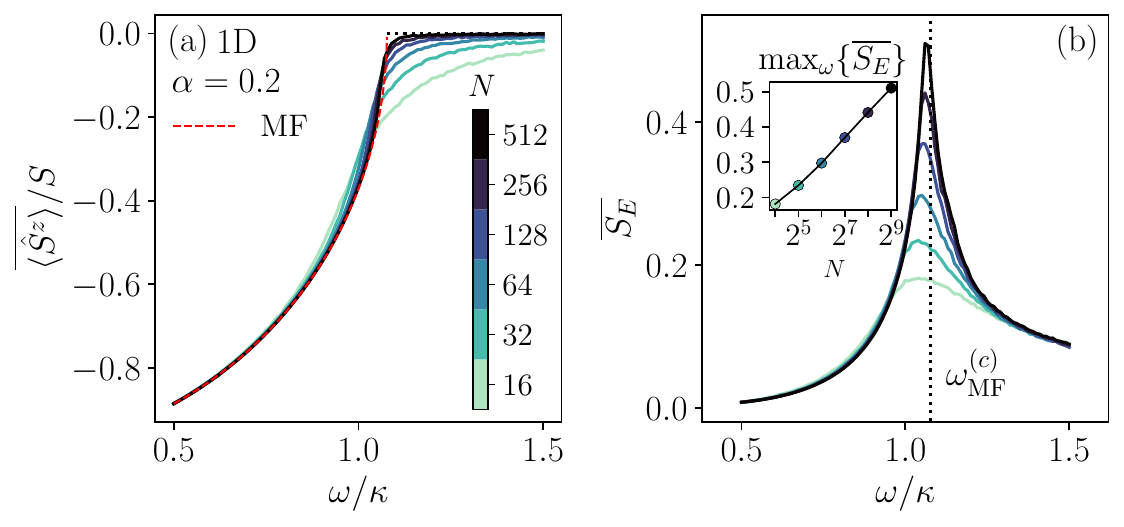}
		\caption{Results for the power-law interacting model in the long-range regime ($\alpha=0.2$, $J=0.1\kappa$) of different system sizes (colorbar shared across panels). The dashed line marks the mean-field solution. (a) Steady-state average $z$ magnetization as a function of the drive $\omega$. (b) Steady-state of the trajectory-averaged half-chain entanglement entropy as a function of $\omega$. The vertical dotted line marks the critical point predicted by the mean-field theory $\omega^{(c)}_{\mathrm{MF}}\simeq 1.077\kappa$. Inset: scaling of the maximum entropy versus $N$ in linear-log scale.}
		\label{fig:btc-zz-long}
	\end{figure}

	    \begin{figure}[h]
    \centering
    \includegraphics[width=\linewidth]{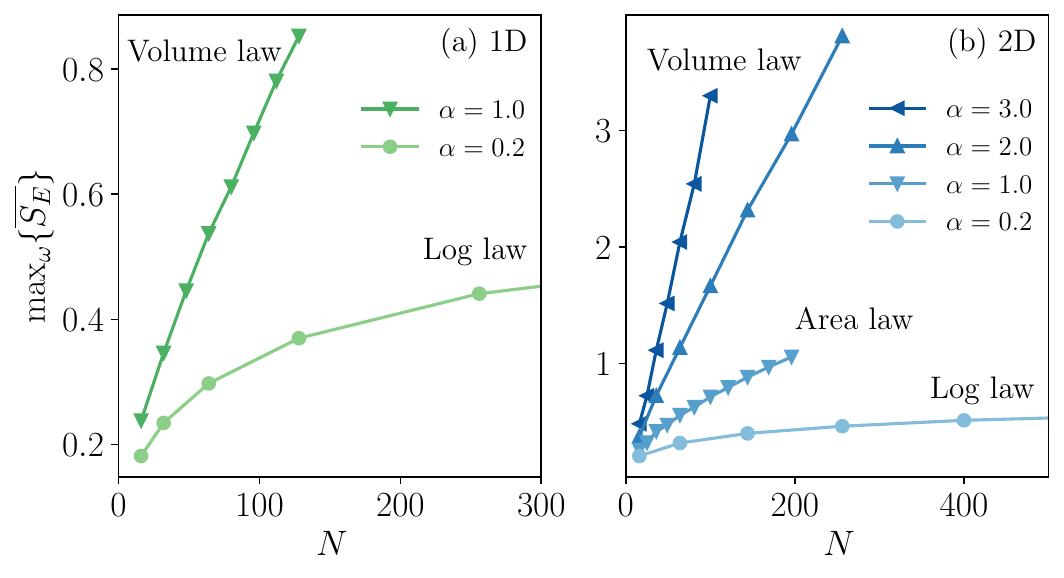}
    \caption{{Maximum steady-state entanglement as a function of system size $N$ (total number of spins) for different interaction range $\alpha$ in (a) 1D and (b) 2D. Different scaling behavior can be observed (see annotation). Note that the horizontal axes are truncated for better visualization and the maximum number of spins we report is $N=1024$ in 2D (see full figures in Supplementary Note~\ref{app:results-2d}).  We report that the value of the drive $\omega$ maximizing the entanglement is asymptotically independent of $N$ (cf. Supplementary Note~\ref{app:results-2d} Fig.~\ref{fig:scaling-argmax}.)}}
    \label{fig:scaling-alpha}
\end{figure}

	\subsubsection{Long-range case: $\alpha \ne 0$}\label{sec:zz-long-range}

	To investigate the effect of a finitely long-range interaction and the entanglement dynamics in this regime, we {first} apply the spin-wave method to the model {in one dimension (1D)} with $\alpha = 0.2$.
Fig.~\ref{fig:btc-zz-long} (a) shows the steady-state expectation of the collective magnetization $\overline{\langle\Shat^z\rangle}$ as a function of the drive amplitude $\omega$ and the system size $N$. As we approach the thermodynamic limit by increasing $N$,  the spin-wave solution converges towards the mean-field prediction and a continuous transition emerges separating the normal phase with $\overline{\langle\Shat^z\rangle}\neq 0$ and the time-crystal phase with $\overline{\langle\Shat^z\rangle}=0$. Fig.~\ref{fig:btc-zz-long} (b) shows the behavior of the long-time-averaged half-chain entanglement entropy across this dissipative phase transition. The entanglement develops a logarithmic divergence as a function of $N$ at the critical point (as implied by the finite-size scaling analysis of the maximum value of the entanglement entropy), while it appears to be $N$-independent for drive values deep inside both phases. This suggests the emergence of an entanglement phase transition separating two area-law phases. We also report (cf. Supplementary Note~\ref{app:power-law-0.2}) that the steady-state spin-wave density $\overline{\epsilon}$ is suppressed with increasing $N$, implying the asymptotic {exactness} of our results. Due to the sufficiently long-range interaction ($\alpha=0.2$), these results bear similar features to the infinite-range model ($\alpha=0$). 
For larger values of $\alpha$, the entanglement dynamics can be dramatically modified, while the dissipative transition witnessed by the magnetization remains however qualitatively similar since the transition is driven by the infinite-range dissipation and not the long-range coherent interaction. We focus therefore on the entanglement in the discussion that follows. As shown in Fig.~\ref{fig:scaling-alpha} (a), the maximum entanglement for $\alpha=1.0$ in 1D exhibits a volume-law scaling in contrast to the log scaling for $\alpha=0.2$ discussed above. (See Supplementary Note~\ref{app:1d-a1} for additional results in this regime.) 

We also perform simulations in 2D (for up to $32\times 32 = 1024$ long-range interacting spins) to study the entanglement scaling as a function of $\alpha$, as shown in Fig.~\ref{fig:scaling-alpha} (b)  for a square lattice of size $N=L^2$. As we shorten the interaction range by increasing $\alpha$, the entanglement scaling changes from the log law to area law ($\propto L$) and volume law ($\propto L^2$), which suggests the presence of a phase transition driven by $\alpha$ between the different entanglement scaling behavior. We report (cf. Supplementary Note~\ref{app:results-2d}) that even in the volume-law phase under moderately short-range settings ($\alpha=2.0$ and $\alpha=3.0$), the spin-wave density remains small at $\overline{\epsilon}\lesssim 0.1$. \hl{On the other hand, the increase of the spin-wave density with higher values of $\alpha$ suggests the eventual breakdown of the spin-wave approximations in the presence of sufficiently short-range interactions.}

	\subsection{Experimental observability of monitored phases boosted by spin-wave quantum trajectories}
	\label{sec:experimental}
	The SWQT framework enables operational advantage in the experimental detection of monitored phases of long-range interacting systems. 
	Compared to Ref.~\cite{Passarelli}, where the post-selection overhead is mitigated by the permutation symmetry of the setup, this section described a method based on quantum-classical correlations~\cite{garratt2023probing,garrattMeasurementsConspireNonlocally2023}.
	This allows to include permutation symmetry-breaking terms, such as local measurements and power-law decaying interactions, of cornerstone importance for current quantum platforms~\cite{Monroe_2021,Ritsch,Lahaye_2009,Bohn_2017,Weimer_2010,Preskill2018quantumcomputingin}.

	In view of the experimental implementations, for example on digital quantum simulators with discrete-time dynamics, we consider a discretized version of the monitored dynamics, where the noise is binned and approximated with a binary random variable, realized via a set of ancilla qubits in the weak measurement formalism (see Sec.~\ref{sec:discrete-dynamics} for a detailed derivation). We consider an experiment with $\mathcal{M}\gg 1$ weak measurement steps (i.e. projective measurements on the ancilla) and set the measurement outcome history $\mvec$. 
	Suppose we perform a final projective measurement of a system operator $\Ohat$ over $\rhohat({\mvec})$, i.e. the state conditioned on the measurement history $\mvec$. This measurement is disruptive and will collapse $\rhohat$ onto the eigenspace of the measurement outcome $o_{\mvec}$. For example, in the case where the observable is chosen to be $\Ohat = \Shat^z$, the final single-shot measurement outcome $o_\mvec\in\{-S, \cdots, S\}$ is one of the $2S+1$ eigenvalues of operator $\Shat^z$.
	Averaging over this quantity recasts the Lindblad prediction, namely $\overline{o_{\mvec}}= \tr[\rhohat\Ohat]$ where $\rhohat=\overline{\rhohat(\mvec)}$ is the average state over all possible trajectories $\mvec$. 
	A correlation that is non-linear in the state can be obtained using classical simulations~\cite{garratt2023probing}. Fixing the measurement history and position $\mvec$ in the classical simulation, we obtain the (classical) estimate of the trajectory-wise expectation value $\langle\Ohat_{\mvec}\rangle_C \equiv \tr[\rhohat_C(\mvec)\Ohat]$ with the label $\bullet_C$ denoting the classically computed quantity. 
	For instance, this is obtained in our setup using the solution of Eq.~\eqref{eq:het-eq-rho} within the spin-wave approximations. We can then cross-correlate the measurement outcome with the classical simulation result to construct the quantum-classical object $o_\mvec\langle\Ohat_{\mvec}\rangle_C$, which, when averaged over $\mvec$ (i.e. over measurement shots), 
	\bea\label{eq:qc-correlator}
	\overline{o_\mvec\langle\Ohat_\mvec\rangle_C} = \overline{\langle\Ohat_\mvec\rangle_Q\langle\Ohat_\mvec\rangle_C}\,,
	\eea
	gives the quantum-classical correlator, where the label $\bullet_Q$ denotes the quantity evaluated on the actual quantum state in the experiment. In the ideal case where the classical simulation is exact, the quantum and classical quantities should coincide, giving a nonlinear function of the state conditioned on the measurement history.
	This was recently shown to reproduce the measurement-induced transition in a variety of experiments~\cite{Google1}, provided simulating $\langle\Ohat_\mvec\rangle_C$ is easy. 
	The SWQT framework therefore is a compelling toolbox for investigating quantum-classical correlations. 
	In particular, when the semi-classical approximation holds, we expect the data to be reliable with errors $O(1/S)$. 
	
	\begin{figure}
		\centering
		\includegraphics[width=0.8\linewidth]{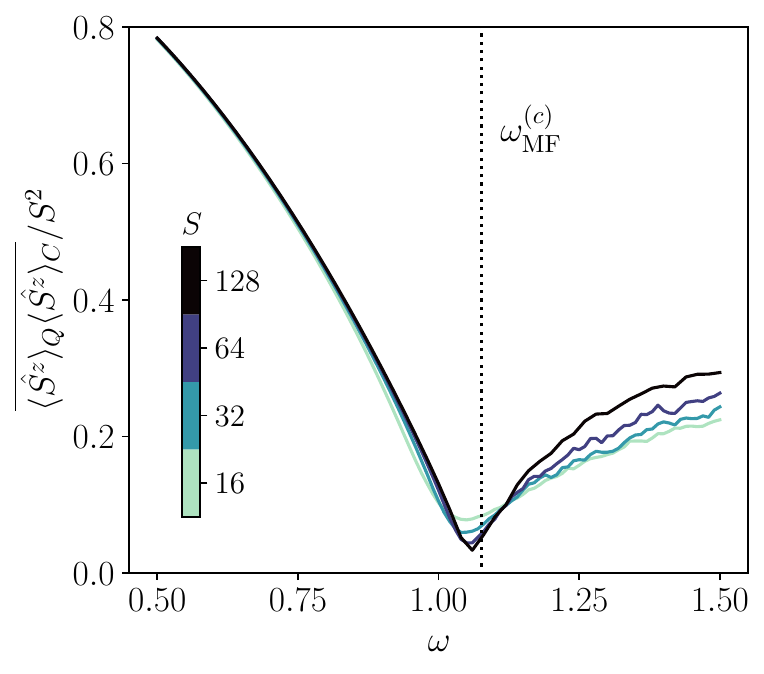}
		\caption{Steady-state trajectory-averaged quantum-classical correlator $\overline{\langle\Shat^z\rangle_Q\langle\Shat^z\rangle_C}$ as a function of the drive $\omega$ for different total spin numbers $S$ (see legend) for the infinite-range model ($\alpha=0$ and $J=0.1\kappa$). The vertical dotted line marks the critical point predicted by the mean-field theory $\omega^{(c)}_{\mathrm{MF}}\simeq 1.077\kappa$. The steady-state value is computed as the average from $\kappa t=50$ to $\kappa t=100$, i.e. over the tail of the dynamics.}
		\label{fig:qc-ss}
	\end{figure}
	
	To demonstrate this quantum-classical measurement protocol, we revisit the infinite-range ($\alpha=0$) case of the power-law spin model (fixing again $J=0.1\kappa$), where the numerically exact solution, that we use to mimic the ``quantum" run, is affordable. The `classical" quantities are then those given by the SWQT method using the same noise realization $\mvec$.
	In Fig.~\ref{fig:qc-ss}, we show the steady-state behavior of the quantum-classical correlator $\overline{\langle\Shat^z\rangle_Q\langle\Shat^z\rangle_C}$  as a function of the drive $\omega$ for different system sizes. Importantly, this quantity successfully captures the signature of the phase transition. This demonstrates that the efficient classical simulation enabled by our spin-wave framework provides access to quantum-classical observables, which can probe nonlinear properties of the state without the post-selection issue.

	\section{Discussion}
	\label{sec:conclusion}

	In this work, we have proposed a stochastic spin-wave theory along quantum trajectories of monitored long-range spin systems. {Our results illustrate several features of the framework that can be summarized as follows:
    \begin{itemize}
        \item The methodology developed here has a wide applicability not only in the study of entanglement phase transitions but can also be exploited to calculate any other quantity, linear and non-linear in the quantum state along the dynamics either on the level of trajectories or on the Lindblad level, by taking trajectory averages.

        \item The proposed method works equally well in any dimension, namely its complexity is always polynomial in the total number of qubits. This overcomes the typical limitation of tensor network methods in two or higher dimensions, including volume-law scaling phases.
    \item When the semi-classical approximations are valid -- including the relevant case of sufficiently long-range interactions, the efficient classical simulation of spin-wave quantum trajectories allows to probe nonlinear quantities of the state via quantum-classical cross-correlated observables, which does not suffer from the post-selection problem, thus opening up promising avenues in the experimental detection of monitored phases in long-range systems.     
    \end{itemize}
    }
    
	Despite the demonstrated performance of our method, there are several interesting directions for future improvements. For example, one can consider terms in the Holstein-Primakoff expansion beyond the leading order to include more nonlinear effects, which should extend the validity of the theory in the regime of relatively high spin-wave densities. We also expect better overall single-trajectory accuracy with the inclusion of higher-order spin-wave corrections, which is essential in improving the performance of the quantum-classical measurement protocol in avoiding the post-selection problem. 
	On the other hand, different unravellings of the Lindblad master equation can also be explored, including the relevant case of quantum jumps modeling the coupling to photodetectors. 

    The long-range spin model we adopted for the demonstration of the spin-wave theory can be readily realized in current experimental platforms~\cite{poggiMeasurementinducedMultipartiteentanglementRegimes2024,ferioliNonequilibriumSuperradiantPhase2023}, and we expect the entanglement phase transition to be readily observable in experiments without the post-selection problem. 
	
	\section{Methods}

	\subsection{Equations of motion for spin-wave quantum trajectories}\label{sec:update-rules}
	
	We provide in this section the technical details for performing the two steps in evolving the variational state ansatz as sketched in Sec.~\ref{sec:swqt}.
	
	\subsubsection{Infinitesimal increments}

	The stochastic master equation~\eqref{eq:het-eq-rho} presented in Sec.~\ref{sec:model} determines the time evolution of the expectation value of operators. For a time-independent operator $\ohat$, the expectation $\langle\ohat\rangle = \tr [\rhohat\ohat ]$ evolves as follows,
	\bea\label{eq:het-eq-dO}
	\d \langle\ohat\rangle = \d t\left\langle \Lcal^\dagger(\ohat)\right\rangle +\sum_i&\[ \d \mbox{w}^*_i\left(\langle\ohat\Lhat_i\rangle -\langle\ohat\rangle\langle \Lhat_i \rangle\right)\right. \\
	& + \left. \d \mbox{w}_i\left(\langle\Lhat^{\dagger}_i\ohat\rangle-\langle \Lhat^{\dagger}_i \rangle\langle\ohat\rangle\right)\]\,,
	\eea
	where $\Lcal^\dagger$ is the adjoint Liouvillian:
	\bea
	\Lcal^\dagger(\ohat)\equiv &~\rmi[\Hhat, \ohat] + \sum_{i,j}f_{ij}\left(\Lhat^\dagger_j\ohat\Lhat_i - \dfrac{1}{2}\left\{\Lhat_j^\dagger\Lhat_i,\ohat\right\}\right)\,.
	\eea
	The infinitesimal increments for the expectation of the time-independent operator $\bbb_i$ can be therefore obtained by setting $\ohat=\bbb_i$ in Eq.~\eqref{eq:het-eq-dO}:
	\bea\label{eq:dbetak}
	\d\beta_i &= \d\langle\bbb_i\rangle\\ &= \rmi\d t \left\langle[ \Hhat,\bbb_i ]\right\rangle\\
	&+ \d t\sum_{j,l}f_{jl}\left\langle\Lhat^\dagger_l\bbb_i\Lhat_j-\dfrac{1}{2}\left\{\Lhat^\dagger_l\Lhat_j,\bbb_i\right\}\right\rangle \\
	&+ \sum_l\(\d \mbox{w}^*_l\langle\delhat_i\Lhat_l\rangle+\d \mbox{w}_l\langle\Lhat^\dagger_l\delhat_i\rangle\)\,.
	\eea
	For the two-point covariance $u_{ij}$, which is the expectation of the time-dependent observable $\delhat_i\delhat_j$, their increments can be obtained using the Ito differentiation rule:
	\bea\label{eq:dukl}
	\d u_{ij} &= \d\langle\delhat_i\delhat_j\rangle\\ &= \rmi\d t \left\langle[ \Hhat,\delhat_i\delhat_j ]\right\rangle\\
	&+ \d t\sum_{l,m}f_{lm}\left\langle\Lhat^\dagger_m\delhat_i\delhat_j\Lhat_l-\dfrac{1}{2}\left\{\Lhat^\dagger_m\Lhat_l,\delhat_i\delhat_j\right\}\right\rangle \\
	&+ \sum_l\[\d \mbox{w}^*_l\left\langle\(\delhat_i\delhat_j-u_{ij}\)\Lhat_l\right\rangle\right.\\
	&\phantom{====}+\left.\d \mbox{w}_l\left\langle\Lhat^\dagger_l\(\delhat_i\delhat_j-u_{ij}\)\right\rangle\]\\
	&-\d t\sum_{l,m}\( f_{lm}\langle\delhat_i\Lhat_l\rangle\langle\Lhat^\dagger_m\delhat_j\rangle+f_{ml}\langle\delhat_j\Lhat_m\rangle\langle\Lhat^\dagger_l\delhat_i\rangle \)\,.
	\eea
	where the last term proportional to $\d t$ comes from the time-dependence of $\delhat_i\delhat_j$.
	Similarly, we have for $v_{ij}$,
	\bea\label{eq:dvkl}
	\d v_{ij} &= \d\langle\ddelhat_i\delhat_j\rangle\\ &= \rmi\d t \left\langle[ \Hhat,\ddelhat_i\delhat_j ]\right\rangle\\
	&+ \d t\sum_{l,m}f_{lm}\left\langle\Lhat^\dagger_m\ddelhat_i\delhat_j\Lhat_l-\dfrac{1}{2}\left\{\Lhat^\dagger_m\Lhat_l,\ddelhat_i\delhat_j\right\}\right\rangle \\
	&+ \sum_l\[\d \mbox{w}^*_l\left\langle\(\ddelhat_i\delhat_j-v_{ij}\)\Lhat_l\right\rangle\right.\\
	&\phantom{====}+\left.\d \mbox{w}_l\left\langle\Lhat^\dagger_l\(\ddelhat_i\delhat_j-v_{ij}\)\right\rangle\]\\
	&-\d t\sum_{l,m}\( f_{lm}\langle\ddelhat_i\Lhat_l\rangle\langle\Lhat^\dagger_m\delhat_j\rangle+f_{ml}\langle\delhat_j\Lhat_m\rangle\langle\Lhat^\dagger_l\ddelhat_i\rangle \)\,.
	\eea
	In the equations above, the Hamiltonian $\Hhat$ and the dissipation operators $\Lhat_i$ should be expressed in terms of the bosonic operators $\bbb_i$ using the substitution rules defined in Eq.~\eqref{eq:bosonization-n}. 
	The Gaussian approximation then allows evaluating the expectation value of every term beyond quadratic order in $\bbb_i$ in terms of one- and two-point correlators, i.e. $\beta_i$, $u_{ij}$ and $v_{ij}$, thanks to the Wick theorem. 
	The increments $\delta\beta_i$, $\delta u_{ij}$, and $\delta v_{ij}$ are then calculated using the discretized versions of the equations above. This can be achieved with the substitution  $\d t \to \delta t$ for a sufficiently small time step $\delta t$. The noise is approximated with $\d \mbox{w}_i \to \delta \mbox{w}_i = \sqrt{\delta t}( X_i + \rmi Y_i)$, where the pair of random vectors $(\mathbf{X},\mathbf{Y})$ at every time step is drawn from a multivariate real Gaussian distribution with zero mean and the following covariance matrix,
	\bea
	\mathbf{K}=\pmx{\mathbf{K}^{\mathbf{X}\mathbf{X}} & \mathbf{K}^{\mathbf{X}\mathbf{Y}} \\ 
		\mathbf{K}^{\mathbf{Y}\mathbf{X}} & \mathbf{K}^{\mathbf{Y}\mathbf{Y}}
	}
	\eea
	with matrix elements 
	\bea
	(\mathbf{K}^{\mathbf{X}\mathbf{X}})_{ij}&=(\mathbf{K}^{\mathbf{Y}\mathbf{Y}})_{ij}=\dfrac{1}{2}\Re f_{ij}\,,\\
	(\mathbf{K}^{\mathbf{X}\mathbf{Y}})_{ij}&=-(\mathbf{K}^{\mathbf{Y}\mathbf{X}})_{ij}=\dfrac{1}{2}\Im f_{ij}\,.
	\eea
	
	We complete this step by updating the Gaussian parameters with the increments obtained following the prescription described above: 
	\bea\label{eq:update-gaussian}
	\beta_i&\leftarrow\beta_i+\delta\beta_i\,,\\
	u_{ij}&\leftarrow u_{ij} + \delta u_{ij}\,,\\
	v_{ij}&\leftarrow v_{ij} + \delta v_{ij}\,.
	\eea
	As a result of the truncated Holstein-Primakoff expansion at the lowest order, the increments include terms up to first order in $1/S$, which account for finite-size effects in the dynamics as a correction to the mean-field (zeroth order) theory.

	\subsubsection{Re-alignment of the frame}\label{sec:realign-frame}
	
	After the infinitesimal evolution of the state, let us update the reference frame such that the $\ztil$ axis aligns with the updated direction of the collective spin. This condition is equivalent to requiring the following quantity to be zero,
	\bea
	\underline{\beta}\equiv \dfrac{1}{N}\sum_{i=1}^N\beta_i\,,
	\eea
	as a direct implication of Eq.~\eqref{eq:bosonization-n}.
	This can be achieved by moving the frame smoothly along a path $\theta(\tau), \phi(\tau)$ parametrized by some parameter $\tau$. The unitary transformation $\Uhat(\theta,\phi)$ therefore becomes $\tau$-dependent,
	\bea
	\uhat(\tau) =  \uhat(\theta(\tau),\phi(\tau))\,.
	\eea
	The rotation of the frame induces some apparent (fictitious) dynamics on the state, whose generator takes the form of an ``inertial Hamiltonian":
	\bea
	\Hhat_\rf &= -\rmi \dfrac{\d \uhat}{\d\tau}\uhat^\dagger\\&= \sin\theta\dfrac{\d \phi}{\d\tau}\Shat^\xtil - \dfrac{\d \theta}{\d\tau}\Shat^\ytil - \cos\theta\dfrac{\d\phi}{\d\tau}\Shat^\ztil\,.
	\eea
	We then use the bosonization rules in Eq.~\eqref{eq:bosonization-n} to substitute the spin operators with bosonic ones to express $\Hhat_\rf$ in terms of $\bbb_i$. To find the amount of rotation of the frame to achieve $\underline{\beta}=0$, let us consider the apparent evolution of the operator $\bbb_i$ along the moving frame:
	\bea\label{eq:beta-align-pde}
	\dfrac{\d\bbb_i}{\d\tau} &= \rmi \left[ \Hhat_\rf, \bbb_i  \right]\\
	&= -\rmi\sqrt{\dfrac{s}{2}}\sin\theta\dfrac{\d\phi}{\d\tau} - \sqrt{\dfrac{s}{2}}\dfrac{\d\theta}{\d\tau}-\rmi\cos\theta\dfrac{\d\phi}{\d\tau}\bbb_i\,.
	\eea    
	This equation can be integrated analytically by considering the path on the Bloch sphere from $(\theta_1,\phi_1)$ to $(\theta_2,\phi_2)$ following the two segments [as illustrated in Fig.~\ref{fig:power-law-spins} (c)]:
	\bea\label{eq:integral-paths}
	\gamma_1 &: \theta(\tau) = \theta_1, \phi(\tau=0) = \phi_1, \phi(\tau=1) = \phi_2 \,;\\
	\gamma_2 &: \phi(\tau) = \phi_2, \theta(\tau=1) = \theta_1, \theta(\tau=2) = \theta_2\,.
	\eea
	The solution is then \bea\label{eq:bn-new}
	\bbb_i(\theta_2,\phi_2) 
	=&~ \[ \sqrt{\dfrac{s}{2}}\tan\theta_1+\bbb_i(\theta_1, \phi_1) \]\rme^{-\rmi\Delta\phi\cos\theta_1}\\ &- \sqrt{\dfrac{s}{2}}\tan\theta_1- \sqrt{\dfrac{s}{2}}\Delta\theta\,,
	\eea
	where $\Delta\theta\equiv\theta_2-\theta_1$ and $\Delta\phi\equiv\phi_2-\phi_1$. 
	This equation immediately implies the evolution of $\underline{\beta}$, upon taking the expectation of both sides,
	\bea
	\underline{\beta}(\theta_2,\phi_2)  =&~ \[ \sqrt{\dfrac{s}{2}}\tan\theta_1+\underline{\beta}(\theta_1, \phi_1) \]\rme^{-\rmi\Delta\phi\cos\theta_1}\\ &- \sqrt{\dfrac{s}{2}}\tan\theta_1- \sqrt{\dfrac{s}{2}}\Delta\theta\,.
	\eea
	As our objective is to find the amount of rotations $\Delta\theta$ and $\Delta\phi$ starting from $\theta_1=\theta$ and $\phi_1=\phi$ such that $\underline{\beta}(\theta+\Delta\theta,\phi+\Delta\phi)=0$, we simply set the right-hand side of the equation above to zero, giving our final expressions for the angle increments:
	\bea\label{eq:angle-increments}
	\Delta\phi &= \dfrac{1}{\cos\theta}\arctan\left\{ \dfrac{ \Im\underline{\beta} }{\sqrt{\frac{s}{2}}\tan\theta + \Re\underline{\beta} }\right\}\\
	\Delta\theta &= \left( \tan\theta+\sqrt{\dfrac{2}{s}}\Re\underline{\beta} \right)\cos(\Delta\phi\cos\theta)\\&+\sqrt{\dfrac{2}{s}}\Im\underline{\beta}\sin(\Delta\phi\cos\theta)-\tan\theta\,.
	\eea
	With the angular increments in hand, the first moments $\beta_i$ of the Gaussian ansatz can be updated directly using Eq.~\eqref{eq:bn-new} with the operator $\bbb_i$ replaced by its expectation value $\beta_i$. This equation also implies the evolution of the fluctuation operators $\delhat_i=\bbb_i-\beta_i$, which is simply
	\bea\label{eq:delhat-rotation}
	\delhat_i(\theta_2,\phi_2)=\delhat_i(\theta_1,\phi_1)\rme^{-\rmi\Delta\phi\cos\theta_1}\,.
	\eea
	We obtain, therefore,
	\bea\label{eq:rotation-beta-u}
	u_{ij}(\theta_2, \phi_2) &= u_{ij}(\theta_1,\phi_1)\rme^{-2\rmi\Delta\phi \cos\theta_1}\,,\\
	v_{ij}(\theta_2,\phi_2) &= v_{ij}(\theta_1, \phi_1)\,.
	\eea
	With this, we complete the full update step, summarized as follows:
	\bea\label{eq:alignment-summary}
	\theta\leftarrow&~\theta+\Delta\theta\,,\\ \phi\leftarrow&~\phi+\Delta\phi\,,\\ \beta_{i}\leftarrow&~ \[ \sqrt{\dfrac{s}{2}}\tan\theta+{\beta_i}\]\rme^{-\rmi\Delta\phi\cos\theta}\\ &- \sqrt{\dfrac{s}{2}}\tan\theta- \sqrt{\dfrac{s}{2}}\Delta\theta\,,\\  u_{ij}\leftarrow&~ u_{ij}\rme^{-2\rmi\Delta\phi \cos\theta}\,,\\ v_{ij}\leftarrow&~ v_{ij} \,.
	\eea
	Finally, we increase the (physical) time $t$ to the next step $t\leftarrow t+\delta t$ and we are ready for a new iteration.
	
	Let us briefly recap the operations performed in each time step of the algorithm. We first update the Gaussian parameters $\beta_i$, $u_{ij}$ and $v_{ij}$ according to Eq.~\eqref{eq:update-gaussian} using the increments computed from Eqs.~\eqref{eq:dbetak}-\eqref{eq:dvkl}. We then perform the re-alignment step which updates the variational parameters according to Eq.~\eqref{eq:alignment-summary} using the angular increments $\Delta\theta$ and $\Delta\phi$ given by Eq.~\eqref{eq:angle-increments}, completing the full iteration.

	\subsection{From continuous to discrete monitored dynamics}\label{sec:discrete-dynamics}
	We discuss in this section the discretized version of the dynamics in Eq.~\eqref{eq:het-eq-rho} by fixing $\Delta t$ a (small) time scale of the unitary and measurement steps, which is equivalent to the continuous-time case in the limit of $\Delta t\to 0$ (see also Refs.~\cite{Brun_2002,Gross_2018,Wiseman}). As shown in Supplementary Note~\ref{app:nondiagonal-unravel}, the stochastic master equation~\eqref{eq:het-eq-rho} can be cast in a diagonal form such that the noises $\d Z_j$ associated with different Lindblad jump operators are independent. We consider therefore the simplified problem with a single Lindblad jump operator $\Lhat$, and focus on the dissipative part of the dynamics (fixing $\Hhat= 0$), namely
	\begin{equation}
		\d\rhohat = \d t\mathcal{D}[\Lhat](\rhohat) + \d Z^* \(\Lhat-\langle \Lhat\rangle\) \rhohat + \d Z \rhohat \(\Lhat^\dagger-\langle \Lhat^\dagger\rangle\)\;,
		\label{eq:reduced}
	\end{equation}
	where the dissipator $\dcal[\Lhat]$ is defined as
	\bea
	\dcal[\Lhat](\rhohat) \equiv \Lhat\rhohat\Lhat^\dagger-\dfrac{1}{2}\left\{\Lhat^\dagger\Lhat,\rhohat\right\}\,,
	\eea
	and the zero-mean noise $\d Z$ satisfies $|\d Z|^2=\d t$ and $\d Z^2=0$.
	The generalization to the complete problem is then trivial.
	
	We discretize also the range of values $\d Z= (\d X + \rmi \d Y)/\sqrt{2}$ into the finite binnings $\d X \in \{ -x_Q,\dots,x_Q\}$ and $\d Y \in \{ - y_Q,\dots,y_Q\}$ for some parameter $Q$. 
	The problem in Eq.~\eqref{eq:reduced} is then the continuous limit of some positive operator-valued measurements (POVM), which can be described using an ancilla $\mathcal{A}$. We fix $\Delta t>0$ a small time-step and study the evolution as a stochastic quantum circuit. 
	Let us consider the dynamics of the state $\hat{\mathfrak{R}}_{\mathcal{S},\mathcal{A}}$ describing the combined system and ancilla framework. 
	By definition, we require that $\rhohat = \mathrm{tr}_\mathcal{A}(\hat{\mathfrak{R}}_{\mathcal{S},\mathcal{A}})$. 
	We note that the choice of ancilla and system interaction is crucial in determining the value of $\d Z$. 
	For simplicity and concreteness, we consider the case of $\mathcal{A}$ being a system of qubits, which has immediate implementations in quantum platforms. In particular, we focus on the minimal setup of two qubits per site, each contributing respectively to the real and imaginary parts of the complex noise. This choice will fix a bimodal approximation of the Gaussian binning, namely $Q=1$. Nevertheless, the argument below can be generalized to more involved ancillas to reproduce larger values of $Q$. 
	
	We denote $\mathcal{A}_{1/2}$ the ancilla qubit 1/2. The heterodyne dynamics in Eq.~\eqref{eq:reduced} is then generated by the system-ancilla interaction
	\begin{equation}
		\begin{split}
			\Uhat_{\mathcal{S},\mathcal{A}}&=  \BS\phantom{.}\rme^{ \sqrt{\Delta t} (\Lhat\otimes \sigmam_1 - \Lhat^\dagger\otimes \sigmap_{1})}\;,\\
			\BS &= \rme^{\pi \left( \sigmap_1\otimes \sigmam_2 - \sigmam_1\otimes \sigmap_2 \right)/4}\,,
		\end{split}
	\end{equation}
	where $\hat{L}$ acts on the system, $\sigmahat^\pm_{1,2}$ are the uppering and lowering operators for the ancilla qubit 1 and 2 respectively, and $\BS$ is the $50/50$ beamsplitter unitary acting only on the ancilla $\mathcal{A}$.

	Applying $\Uhat_{\mathcal{S},\mathcal{A}}$ to $\hat{\mathfrak{R}}_{\mathcal{S},\mathcal{A}}(t)\equiv \rhohat(t) \otimes |00\rangle\langle 00|$ (with the convention where $|0\rangle$ denotes the spin-up state) and projecting out $\mathcal{A}_1$ onto the basis $|\pm\rangle \equiv (|0\rangle \pm |1\rangle)/\sqrt{2}$ and $\mathcal{A}_2$ onto $|\tilde{\pm}\rangle \equiv (|0\rangle \tilde{\pm} \rmi |1\rangle )/\sqrt{2}$ we have the four Kraus operators
	\begin{equation}
		\begin{split}
			\Khat_{\pm,\tilde{\pm}}  &\equiv \langle \pm |_{\mathcal{A}_1}\langle \tilde{\pm}|_{\mathcal{A}_2} \Uhat_{\mathcal{S},\mathcal{A}} |0\rangle_{\mathcal{A}_1}|0\rangle_{\mathcal{A}_2}\;\\
			&= \frac{1}{2}\left(\mathbb{I} \pm e^{\mp \tilde{\pm} \rmi \pi/4}\sqrt{\Delta t} \Lhat - \frac{1}{2} \Delta t \Lhat^\dagger \Lhat\right)\;,
		\end{split}
	\end{equation}
	where we have kept terms up to first order in $\Delta t$.
	We note that the measurement information in the Kraus operators is fully encoded in the 4 complex numbers $\pm e^{\mp\tilde{\pm} \rmi \pi/4}$. Their choice, fixing the measurement history, is therefore determined by the POVM $\Ehat_{\pm,\tilde{\pm}} = \Khat_{\pm,\tilde{\pm}}^\dagger \Khat_{\pm,\tilde{\pm}}$.
	The post-measurement state is 
	\begin{equation}
		\rhohat_{\pm,\tilde{\pm}} = \dfrac{\Khat_{\pm,\tilde{\pm}}\rhohat \Khat^\dagger_{\pm,\tilde{\pm}}}{p_{\pm,\tilde{\pm}}}\,,
		\label{eq:test}
	\end{equation}
	where the probabilities are given by
	\bea
	p_{\pm,\tilde{\pm}} = \mathrm{tr}(\rhohat \Ehat_{\pm,\tilde{\pm}})\,.
	\eea
	We will now briefly show that Eq.~\eqref{eq:test} reproduces Eq.~\eqref{eq:reduced} for small $\Delta t$, with a binary approximation of the binning. 
	We define two stochastic variables depending on the measurement outcomes $\Delta R_{x}(\pm,\tilde{\pm}) \equiv \pm \sqrt{\Delta t}$ and $\Delta R_{y}(\pm,\tilde{\pm}) \equiv \tilde{\pm} \sqrt{\Delta t}$. We then have, at leading order in $\Delta t$, that 
	\begin{equation}
		\begin{split}
			\overline{\Delta R_x} &= \sum_{\pm,\tilde{\pm}} \pm \sqrt{\Delta t} p_{\pm,\tilde{\pm}} = \Delta t \langle \Lhat^\dagger + \Lhat\rangle/\sqrt{2}\,,\\
			\overline{\Delta R_y} &= \sum_{\pm,\tilde{\pm}} \tilde{\pm} \sqrt{\Delta t} p_{\pm,\tilde{\pm}} = \rmi\Delta t \langle \Lhat^\dagger - \Lhat\rangle/\sqrt{2}\,.
		\end{split}
	\end{equation}
	In a similar fashion, we have $\overline{\Delta R_x^2} = \overline{\Delta R_y^2} = \Delta t$ and $\overline{\Delta R_x \Delta R_y} = 0$. Putting all together, and defining the zero-mean stochastic processes $\d X \equiv \Delta R_x - \overline{\Delta R_x}$ and $\d Y = \Delta R_y - \overline{\Delta R_y}$, which satisfy $\overline{\d X^2}=\overline{\d Y^2} = \Delta t $ and $\overline{\d X \d Y}=0$, we obtain the final expression after simple algebra,
	\begin{equation}
		\begin{split}
			\delta\rhohat_{\pm,\tilde{\pm}} &= \rhohat_{\pm,\tilde{\pm}}- \rhohat \\
			&= \Delta t \mathcal{D}[\Lhat]\rhohat + \frac{1}{\sqrt{2}}\d X \(\Lhat\rhohat + \rhohat \Lhat^\dagger - \langle \Lhat + \Lhat^\dagger\rangle\rhohat\) \\
			&\quad + \frac{\rmi}{\sqrt{2}}\d Y \(-\Lhat\rhohat + \rhohat \Lhat^\dagger +\langle \Lhat - \Lhat^\dagger\rangle\rhohat\)\;,
		\end{split}
	\end{equation}
	which is equivalent to Eq.~\eqref{eq:reduced} when substituting $\d Z = (\d X+\rmi\d Y)/\sqrt{2}$. \hl{We benchmark the validity of this binary approximation in Supplementary Note~\ref{sec:bin-benchmark}.}
	
	In summary, within the choice of ancilla and system-ancilla interaction, we obtain a 4-valued complex process $\d Z$ for each independent Lindblad jump operator $\Lhat$ at each monitoring step. This allows us to estimate the brute-force post-selection overhead over $M$ discrete timesteps for $\Lambda$ independent Lindblad jump operators as $O(4^{M\Lambda})$. 
	More generally, introducing further ancilla qubits, or enabling qubit interactions, allows one to reach a more refined binning of the Gaussian increment $\d Z$. Fixing the binning parameter $Q$, the probability of reproducing the same trajectory is then $O((2Q)^{2M\Lambda })$.
	This renders the brute-force experimental observation of the monitored phases of generic systems an exponentially hard task.

	\section*{Data availability}
    The data generated in this study has been deposited in the Zenodo public folder~\cite{turkeshiDataMonitoredLongrange2025}.

	\section*{Code availability}
	The computer code developed in this study is available from the authors upon reasonable request. A demo is also provided in the Code Ocean capsule~\cite{dc0846b7-ce64-47b7-bc0c-521f8994765a}.
	
	\section*{Acknowledgments}
	We would like to thank Marco Schir\`o, Gerald E. Fux, Procolo Lucignano, Gianluca Passarelli, Silvia Pappalardi, Cristiano Ciuti, and Valentin Heyraud for helpful discussions. This work was supported by PNRR MUR project PE0000023- NQSTI, by the European Union (ERC, RAVE, 101053159). X.T. acknowledges support from DFG under Germany's Excellence Strategy – Cluster of Excellence Matter and Light for Quantum Computing (ML4Q) EXC 2004/1 – 390534769, and DFG Collaborative Research Center (CRC) 183 Project No. 277101999 - project B01. Views and opinions expressed are however those of the author(s) only and do not necessarily reflect those of the European Union or the European Research Council. Neither the European  Union nor the granting authority can be held responsible for them.

    \section*{Author contributions}
    Z.L., A.D., X.T. and R.F. contributed to the conception and the implementation of the research, to the discussion of the results, and to the writing of the manuscript.

    \section*{Competing interests}
    The authors declare no competing interests.
	
	\bibliography{BiblioEPT_Semi}
	
	% \clearpage
	
%%%%%%%%%% Merge with supplemental materials %%%%%%%%%%
\widetext
\clearpage
\begin{center}
\textbf{\large Supplementary Information for \textit{Monitored long-range interacting systems: spin-wave theory for quantum trajectories}}
\end{center}
%%%%%%%%%% Merge with supplemental materials %%%%%%%%%%
%%%%%%%%%% Prefix a "S" to all equations, figures, tables and reset the counter %%%%%%%%%%
\setcounter{equation}{0}
\setcounter{figure}{0}
\setcounter{table}{0}
\setcounter{page}{1}
\makeatletter
\renewcommand{\theequation}{S\arabic{equation}}
\renewcommand{\thefigure}{S\arabic{figure}}
% \renewcommand{\bibnumfmt}[1]{[S#1]}
% \renewcommand{\citenumfont}[1]{S#1}
%%%%%%%%%% Prefix a "S" to all equations, figures, tables and reset the counter %%%%%%%%%%

\appendix
\twocolumngrid

\graphicspath{{./}{./fig_supp/}}
	
	\section{Proof of the unraveling of the nondiagonal Lindblad master equation}\label{app:nondiagonal-unravel}
	
	In the nondiagonal master equation~\eqref{eq:swqt-master-eq} of the main text,
	the positive semidefinite matrix $f_{ij}$ can be diagonalized with a unitary transformation $u$:
	\bea
	f = u\kappa u^\dagger\,,
	\eea
	where $\kappa$ is a diagonal matrix with non-negative entries:
	\bea
	\kappa\equiv\mathrm{diag}(\kappa_1,\ldots,\kappa_N)\,.
	\eea
	
	Defining 
	\bea
	\Ahat_i&\equiv\sum_j u_{ji}\Lhat_j\,,\\
	\d Z_i&\equiv\dfrac{1}{\sqrt{\kappa_i}}\sum_{j}u_{ji}\d \mbox{w}_j\,,
	\eea
	the Liouvillian~\eqref{eq:swqt-master-eq} in the main text becomes
	\bea
	\Lcal(\rhohat)=-\rmi[\Hhat, \rhohat]+\sum_i\kappa_i\dcal[\Ahat_i](\rhohat)\,,
	\eea
	where 
	\bea
	\dcal[\Ahat_i](\rhohat)\equiv\Ahat_i\rhohat\Ahat_i^\dagger - \dfrac{1}{2}\left\{\Ahat_i^\dagger\Ahat_i,\rhohat\right\}
	\eea
	is the standard (diagonal) dissipator.
	The unraveling~\eqref{eq:het-eq-rho} in the main text becomes
	\bea\label{eq:het-diag-eq-rho}
	\d\hat\rho &= \dt\Lcal(\rhohat) \\
	&+ \sum_i\sqrt{\kappa_i}\[ \d Z^*_i\left(\Ahat_j-\langle \Ahat_i \rangle\right)\rhohat  + \d Z_i\rhohat\left(\Ahat^{\dagger}_i-\langle \Ahat^{\dagger}_i \rangle\right)\]\,,
	\eea
	where the noise satisfies
	\bea
	&\overline{\d Z_i}=0\,,&\\
	&\d Z^*_i\d Z_j=\delta_{ij}\d t\,,&\d Z_i\d Z_j=0\,,
	\eea
	which implies that each $\d Z_i$ is a normalized complex Wiener process and independent from each other. Eq.~\eqref{eq:het-diag-eq-rho}, being equivalent to Eq.~\eqref{eq:het-eq-rho} of the main text, is the standard (diagonal-form) quantum state diffusion unraveling describing a heterodyne detection process, where the operators $\Ahat_i$ are being continuously monitored at rates $\kappa_i$ respectively.
	
	For a pure initial state, this is equivalent to the following stochastic \schr{} equation,
	
	\bea
	&\d|\psi\rangle = -\rmi\Hhat\dt|\psi\rangle\\
	&+ \sum_i\kappa_i\(\left\langle \Ahat^\dagger_i \right\rangle\Ahat_i - \dfrac{1}{2}\left\langle\Ahat^\dagger_i\right\rangle\left\langle\Ahat_i\right\rangle-\dfrac{1}{2}\Ahat^\dagger_i\Ahat_i\)\dt|\psi\rangle\\
	&+\sum_i\sqrt{\kappa_i}\(\Ahat_i-\left\langle\Ahat_i\right\rangle\)\d Z^*_i|\psi\rangle\,,
	\eea
	which preserves the purity of the state along every single trajectory.
	
	\section{Mean-field equations for the power-law spin model}\label{app:mf-power-law-spin}
	We derive in this Section the mean-field equations of the power-law spin model defined in Sec.~\ref{sec:power-law-spin-model} of the main text. We denote the average magnetization with
	\bea
	m_{\mu}\equiv\dfrac{\langle\Shat^{\mu}\rangle}{S}=\dfrac{1}{N}\sum_{i=1}^N\langle\sigmahat^\mu_i\rangle\,,\quad \mu\in\{x,y,z\}\,,
	\eea
	whose evolution under the Lindblad dynamics can be obtained with the adjoint master equation~\cite{breuer2002theory}:  for a time-independent operator $\ohat$, we have
	\bea
	\dfrac{\d\langle\ohat\rangle}{\d t} ={}&\rmi \left\langle[ \Hhat,\ohat ]\right\rangle + \dfrac{\kappa}{S}\left\langle\Shat^+\ohat\Shat^--\dfrac{1}{2}\left\{\Shat^+\Shat^-,\ohat\right\}\right\rangle\,.
	\eea
	Within the mean-field approximation, where we assume the factorization $\langle\sigmahat_i^\mu\sigmahat_j^\lambda\rangle=\langle\sigmahat_i^\mu\rangle\langle\sigmahat^\lambda_j\rangle$ at $N\to\infty$, and that $\langle\sigmahat^\mu_i\rangle$ has no $i-$dependence, we obtain the following equations of motion for the average magnetization:
	\bea
	\dfrac{\d m_x}{\d t} &= -4J m_y m_z + \kappa m_x m_z\,,\\
	\dfrac{\d m_y}{\d t} &= -\omega m_z+4J m_x m_z + \kappa m_y m_z\,,\\
	\dfrac{\d m_z}{\d t} &= \omega m_y - \kappa(m_x^2+m_y^2)\,.
	\eea
	The steady-state magnetization can be found by imposing the time derivatives to zero, which yields
	\bea
	m_z = -\sqrt{1-\dfrac{\omega^2}{16J^2+\kappa^2}}\,.
	\eea
	Beyond the critical point $\omega_\mathrm{MF}^{(c)} = \sqrt{16J^2+\kappa^2}$, this stationary solution no longer exists and the magnetization admits only permanently oscillating solutions with zero mean when averaged over long times. The mean-field theory therefore predicts a continuous phase transition from a stationary phase to a time-crystal phase on the level of the Lindblad dynamics of the average state.

	\section{Expression for the entanglement entropy under the Gaussian approximation}\label{app:gaussian-entanglement}
	
	Without loss of generality, let us consider a bipartition of the spins where one subsystem contains spins indexed from 1 to $M$. Denoting $\xhat_i\equiv(\dbbb_i+\bbb_i)/\sqrt{2}$, $\phat_i\equiv\rmi(\dbbb_i-\bbb_i)/\sqrt{2}$ and $\hat{\rvec}\equiv(\xhat_1,\ldots,\xhat_M, \phat_1,\ldots,\phat_M)$, the covariance matrix $\mathbf{\Xi}$ for the Gaussian state of the subsystem has components
	\bea
	\Xi_{ij}\equiv\frac{1}{2}\langle  \hat{r}_i \hat{r}_j +  \hat{r}_j \hat{r}_i \rangle - \langle\hat{r}_i\rangle\langle\hat{r}_j\rangle\,,
	\eea 
	which can be expressed in terms of $u_{ij}$ and $v_{ij}$.
	The entanglement entropy $S_E$ is then given by the von Neumann entropy of the subsystem (since the full system remains in a pure state along the unraveled trajectory), which can be expressed in terms of the symplectic eigenvalues $\{\mu_i\}_i$ of the covariance matrix $\mathbf{\Xi}$ as~\cite{serafiniSymplecticInvariantsEntropic2003}
	\bea
	S_E &=\sum_i\[\left(\mu_i + \dfrac{1}{2}\right) \log\left(\mu_i + \dfrac{1}{2}\right)\right.\\
	&\phantom{===}\left.- \left(\mu_i - \dfrac{1}{2}\right)  \log\left(\mu_i - \dfrac{1}{2}\right)\right]\,.
	\eea    
	Operationally, the symplectic eigenvalues can be found with the help of the symplectic matrix $\mathbf{\Omega}$ defined as
	\bea
	\mathbf{\Omega} = \pmx{
		\mathbf{0} & \mathbf{I}_M \\
		-\mathbf{I}_M & \mathbf{0}
	}\,,
	\eea
	where $\mathbf{I}_M$ is the $M\times M$ identity matrix. The eigenvalues of the matrix $\mathbf{\Xi}\mathbf{\Omega}$ are then $\{\pm\rmi \mu_i\}_i$, where the $\mu_i$'s are the symplectic eigenvalues of $\mathbf{\Xi}$.

	\section{Additional numerical results}
	
	\subsection{Benchmark of SWQT on the steady-state of the infinite-range model}\label{app:benchmar-steady-state}

	\begin{figure}
		\centering
		\includegraphics[width=0.9\linewidth]{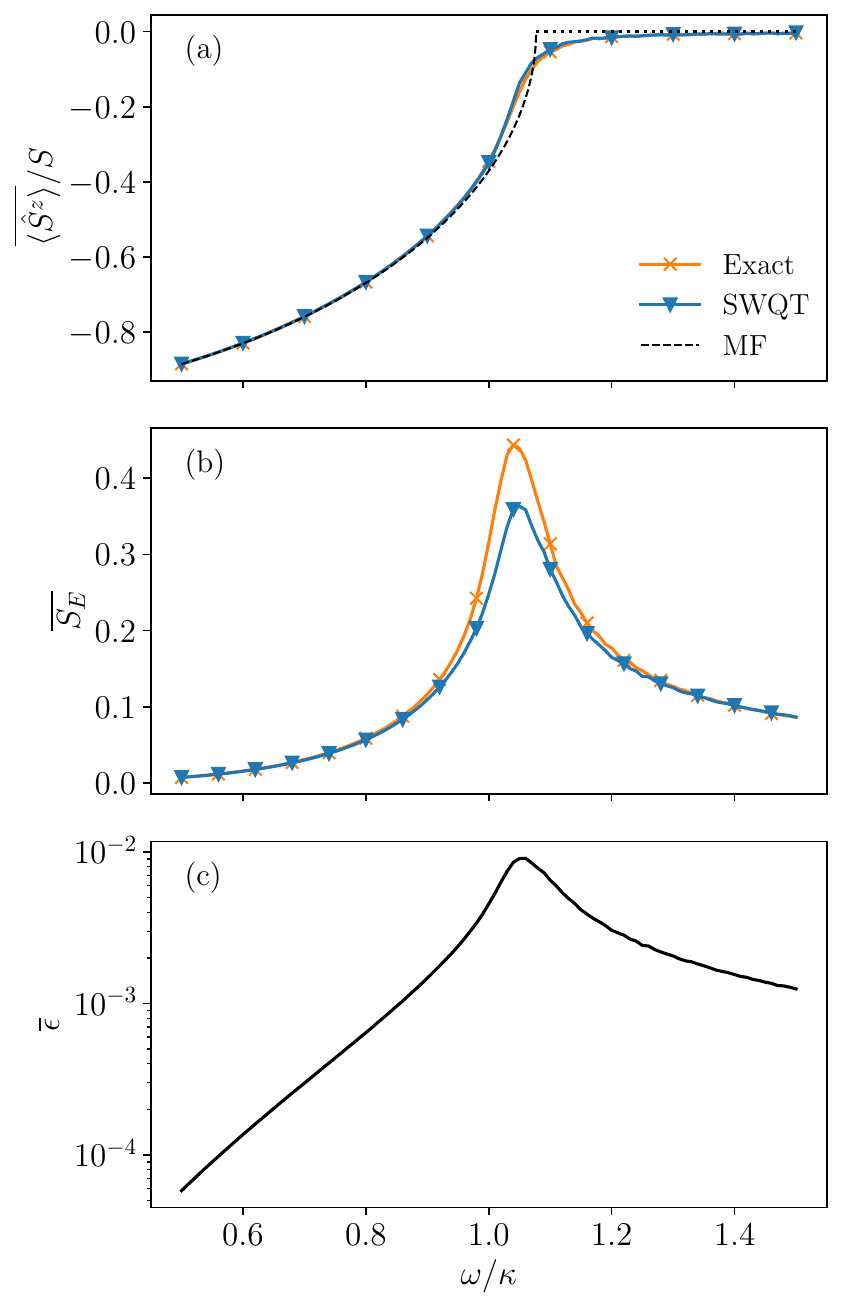}
		\caption{Benchmark of the spin-wave method on the steady state of the collective model with  $S=64$ and $J=0.1\kappa$. (a) Expectation of the collective $z$-magnetization as a function of the drive $\omega$ obtained with the spin-wave quantum trajectories and the exact solution (see legend). The mean-field solution (see supplementary) is marked with the dashed line. (b) Trajectory-averaged steady-state half-chain entanglement entropy obtained with the two methods [see legend in (a)].  (c) Trajectory-averaged spin-wave density $\overline{\epsilon}$ in the steady state.  }
		\label{fig:bench-collective-ss}
	\end{figure}
	
	In Fig.~\ref{fig:bench-collective-ss}, we compare the steady-state of trajectory average quantities from the spin-wave solutions on the infinite-range model as considered in Sec.~\ref{sec:infinite-range} of the main text, against the exact ones, for a wide range of the drive $\omega$. Panel (a) shows the magnetization and panel (b) shows the half-system entanglement entropy. Panel  (c) shows the spin-wave density $\overline{\epsilon}$ associated with the spin-wave solutions. The peak in $\overline{\epsilon}$ corresponds to the maximum in the entanglement entropy [in panel (b)], where the difference between the spin-wave and exact solutions is also more pronounced. This illustrates the significance of the control parameter $\overline{\epsilon}$ as a signal for the validity of the spin-wave theory. On the other hand, the qualitative behavior of the entanglement is correctly captured despite the relatively high spin-wave density at its peak, and the magnetization predicted by the spin-wave method [panel (a)] remains accurate for all values of $\omega$ considered.
	
	\subsection{Comparison with other existing spin-wave methods}\label{app:swqt-vs-dsw}
	
	\begin{figure*}[ht]
		\centering
		\includegraphics[width=0.9\linewidth]{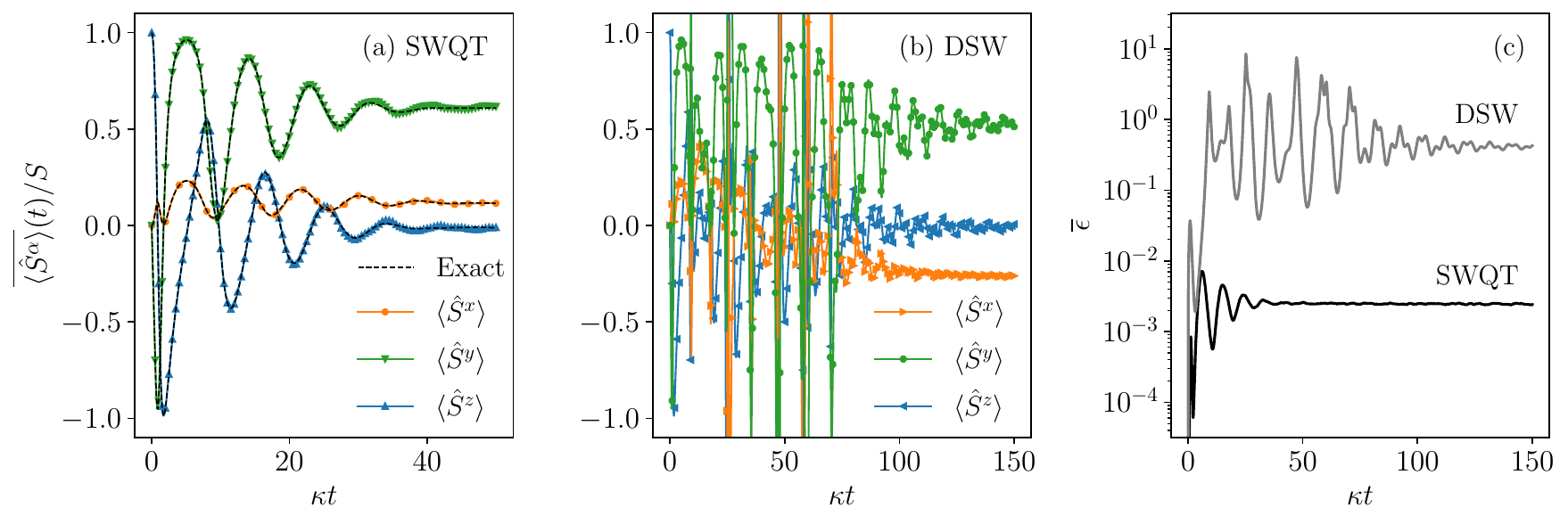}
		\caption{Comparison between solutions given by the two different spin-wave methods. (a) Results from the spin-wave quantum trajectories (SWQT) on the expectation value of the collective spin vector in the time-crystal regime with parameters $\omega_F=1.25\kappa$, $J=0.1\kappa$ and  $S=64$. (b) The solution given by the deterministic spin-wave theory (DSW) with the same parameters. (c) Spin-wave densities associated with the two solutions.}
		\label{fig:benchmark-swqt-vs-dsw}
	\end{figure*}

	In this section, we provide a comparison between our method of spin-wave quantum trajectories (SWQT) and the method proposed in~\cite{seetharamDynamicalScalingCorrelations2022}, which is a deterministic spin-wave method for dissipative systems (referred to as ``DSW" in this section). The latter assumes the same approximations as in our method (i.e. lowest-order Holstein-Primakoff expansion and Gaussian approximation) except that their approximations are performed on the level of the averaged state (i.e. the density matrix) and their dynamical evolution is derived from the deterministic Lindblad master equation. 
	
	For illustration purposes, let us consider again the collective spin model defined in Sec.~\ref{sec:infinite-range} of the main text. The solutions given by the two methods are shown in Fig.~\ref{fig:benchmark-swqt-vs-dsw} for comparison. Panel (a) shows the SWQT results for the parameters $\omega_F=1.25\kappa$, $J=0.1\kappa$,  $S=64$. The solution with the same parameters given by DSW is shown in panel (b), which is by no means close to the exact solution. As explained in the main text, this is a result of the highly mixed nature of the average state that the spin-wave approximations fail to capture. Panel (c) shows the spin-wave densities associated with the two solutions, where the SWQT has a spin-wave density that is smaller by several orders of magnitude, which is consistent with its accuracy compared to the exact solution.

    \subsection{Additional results for the power-law spin model with $\alpha=0.2$ {in 1D}}\label{app:power-law-0.2}

	\begin{figure}[t]
		\centering
		\includegraphics[width=\linewidth]{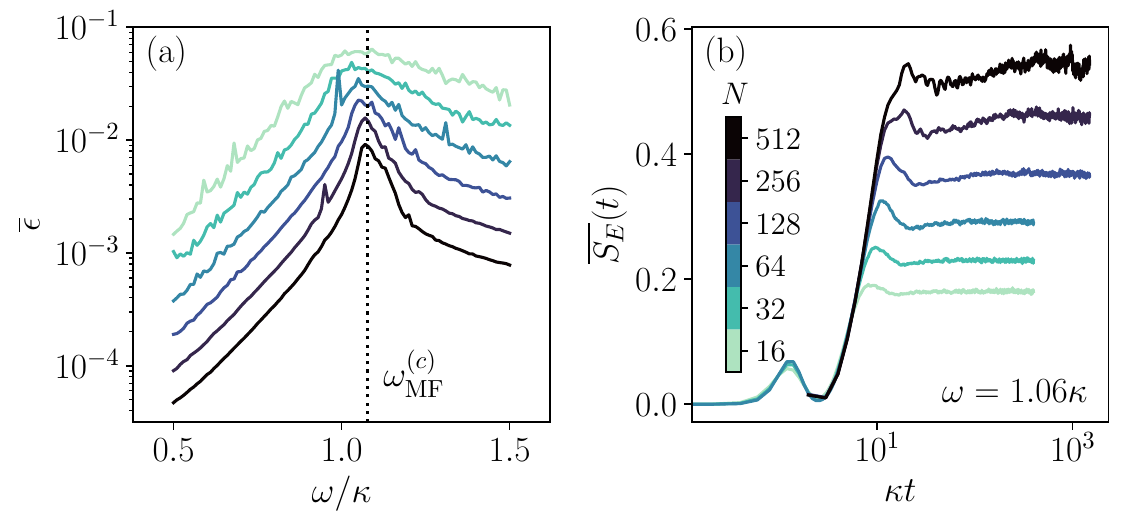}
		\caption{Results for the power-law interacting model in the {case of} $\alpha=0.2$, $J=0.1\kappa$ of different system sizes {in 1D} (colorbar shared across panels). (a) The scaling of the steady-state average spin-wave density $\overline{\epsilon}$ as a function of the drive $\omega$, in log-linear scale. The vertical dotted lines mark the critical point predicted by the mean-field theory $\omega^{(c)}_{\mathrm{MF}}\simeq 1.077\kappa$. (b) Time evolution of the trajectory-average entanglement entropy at a driving value close to criticality $\omega=1.06\kappa$, in linear-log scale. }
		\label{fig:btc-zz-long-st-nsw}
	\end{figure}

	\begin{figure*}
		\centering
		\includegraphics[width=\linewidth]{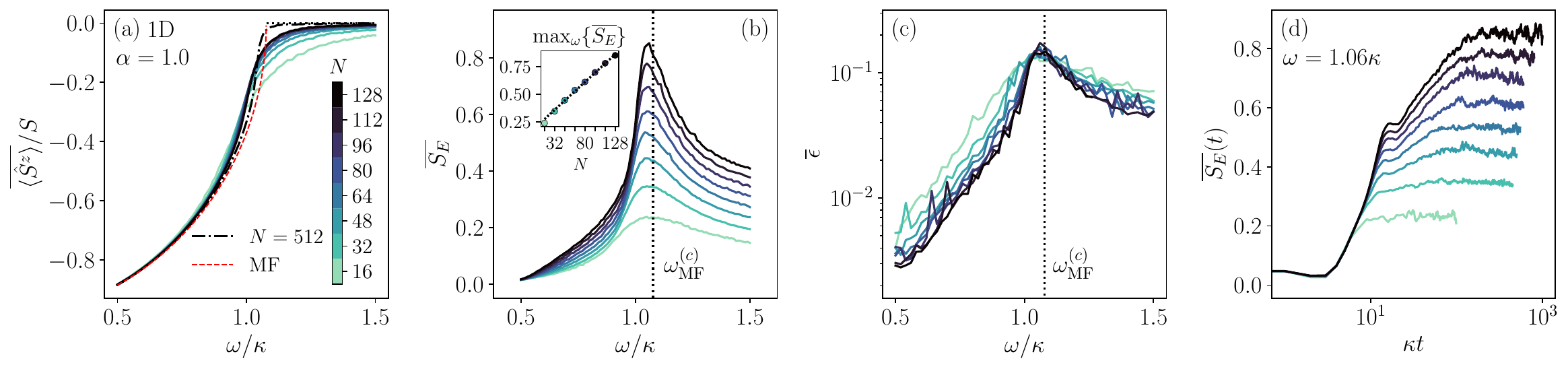}
		\caption{Results for the power-law interacting model {in the case of} $\alpha=1.0$, $J=0.1\kappa$ {for} different system sizes {in 1D} (colorbar shared across all panels). (a) Steady-state average $z$ magnetization as a function of the drive $\omega$. The dashed line marks the mean-field solution.  (b) Steady-state of the trajectory-averaged half-chain entanglement entropy as a function of $\omega$. The vertical dotted lines mark the critical point predicted by the mean-field theory $\omega^{(c)}_{\mathrm{MF}}\simeq 1.077\kappa$. Inset: scaling of the maximum entropy versus $N$ in linear scale. \hl{The dotted line is a linear fit for $N$ ranging from 32 to 128 and we report the coefficient of determination to be $R^2=0.9980$.} (c) Steady-state spin-wave density in log-linear scale. (d) Dynamics of the trajectory-averaged half-chain entanglement entropy for a driving value $\omega=1.06\kappa$, in linear-log scale.}
		\label{fig:btc-zz-short}
	\end{figure*}

	We study the scaling of the steady-state spin-wave density of the power-law spin model {on a 1D chain} with $\alpha=0.2$ as considered in Sec.~\ref{sec:zz-long-range} of the main text, as shown in Fig.~\ref{fig:btc-zz-long-st-nsw} (a). When the system size $N$ increases, the spin-wave density is suppressed (including its peak value), which is a signature of the long-range nature of the model: in the $N\to\infty$ limit, a vanishing spin-wave density suggests that the system becomes equivalent to a mean-field (infinite-range) one. In Fig.~\ref{fig:btc-zz-long-st-nsw} (b), we study the time evolution of the trajectory-averaged entanglement entropy 
	at a driving value $\omega=1.06\kappa$, which is close to where the maximum steady-state entanglement is achieved for the finite system sizes we studied. Note that we are increasing $N$ exponentially between the different considered values, while the entanglement quickly reaches the steady-state value in all cases, suggesting the mitigation of the exponential overhead for observing the entanglement transition via brute-force post selection.

	\subsection{Results for the power-law spin model with $\alpha=1$ {in 1D}}\label{app:1d-a1}

	To study the effect of the interaction range on the entanglement dynamics in the power-law spin model, we also consider the shorter-range case with $\alpha=1$ {on a 1D chain} for comparison. The steady-state magnetization $\overline{\langle\hat{S}^z\rangle}$, as shown in Fig.~\ref{fig:btc-zz-short} (a), exhibits qualitatively similar behavior to the long-range case, while it shows more deviation from the mean-field prediction. This could result from more significant finite-size effects due to short-range interactions. (Note that the mean-field theory does not account for the interaction range parameter $\alpha$ and therefore gives identical predictions regardless of $\alpha$.) In sharp contrast, Fig.~\ref{fig:btc-zz-short} (b) shows that the entanglement entropy of the short-range systems grows much faster with the system size $N$. Finite-size scaling suggests a volume law for the maximum entanglement entropy as well as in a vicinity around the critical point. The spin-wave density, as shown in Fig.~\ref{fig:btc-zz-short} (c), also presents a qualitatively different behavior from the long-range case. Close to the maximal entanglement, the spin-wave density does not appear to decrease with the system size $N$, and remains around $\overline{\epsilon}\lesssim 0.2$. This suggests that the short-range model does not reduce to a mean-field (infinite-range) one even in the thermodynamic limit of $N\to\infty$, and that higher-order corrections to the theory are expected to have a more significant contribution.  Finally, we show in Fig.~\ref{fig:btc-zz-short} (d) the time dynamics of the trajectory-averaged entanglement entropy at $\omega=1.06\kappa$. Contrary to the fast saturation of entanglement in the long-range (or infinite-range) regime, the time it takes to reach the steady-state value scales rapidly with $N$.  {Despite} the relatively high spin-wave density at the considered driving, we expect our result to hold qualitatively.

    \begin{figure*}[h]
        \centering
        \includegraphics[width=\linewidth]{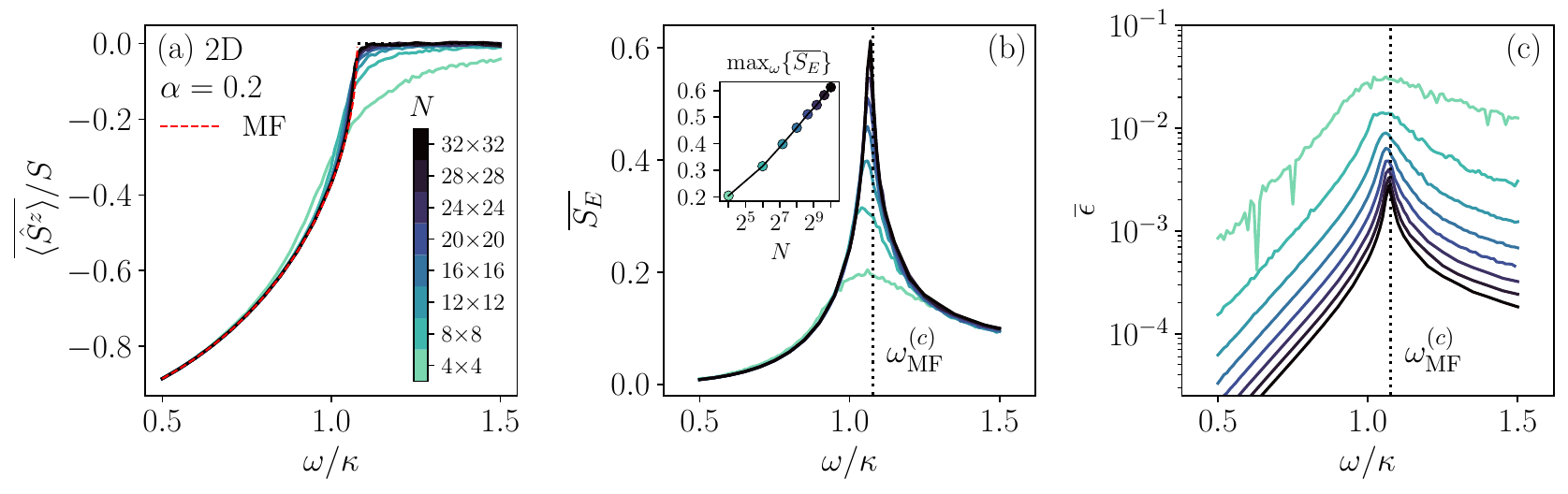}
        \caption{{Results for the power-law spin model in 2D with $\alpha=0.2$, $J=0.1\kappa$ for the following trajectory-averaged quantities in the steady state: (a) $z$ magnetization $\overline{\langle\hat{S}^z\rangle}/S$ with the mean-field solution marked by the dashed line, (b) half-system entanglement entropy $\overline{S_E}$ with its maximum value in the inset, and (c) spin-wave density $\overline{\epsilon}$, as functions of the drive $\omega$ for different system sizes $N=L^2$ (see colorbar).} }
        \label{fig:btc-zz-a02}
    \end{figure*}
    
    \begin{figure*}[h]
        \centering
        \includegraphics[width=\linewidth]{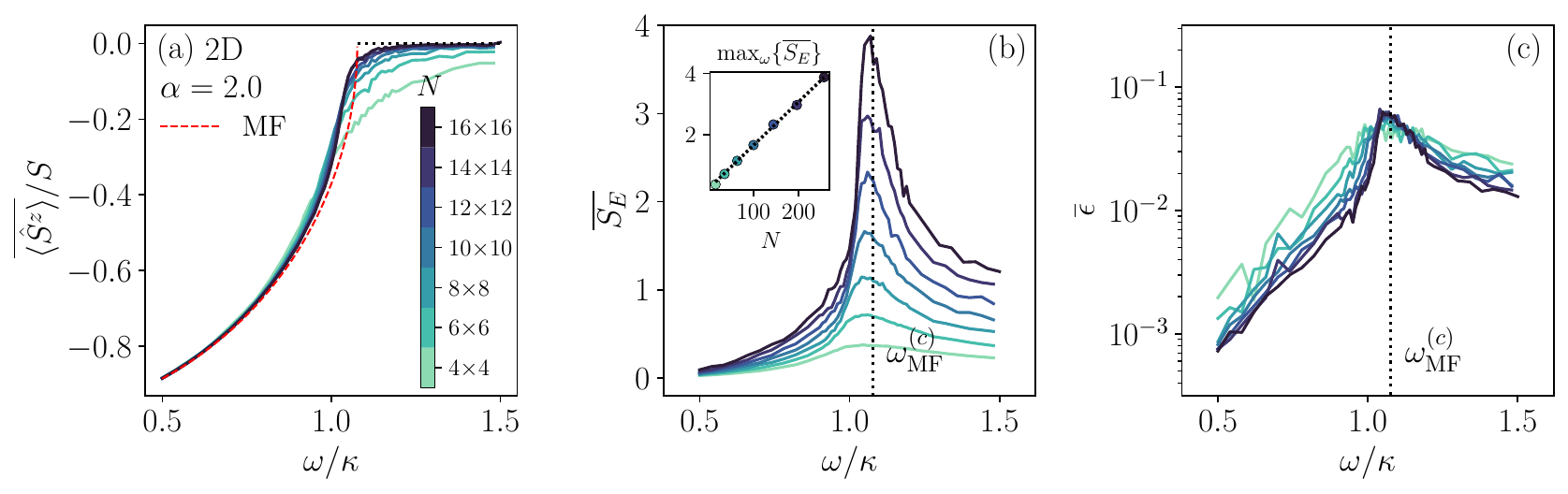}
        \caption{{Same as Fig.~\ref{fig:btc-zz-a02} but for $\alpha=2.0\,$.} \hl{The dotted line in the inset is a linear fit for $N$ ranging from 36 to 256 and we report the coefficient of determination to be $R^2=0.9991$.}}
        \label{fig:btc-zz-a2}
    \end{figure*}
    
    \begin{figure*}[h]
        \centering
        \includegraphics[width=\linewidth]{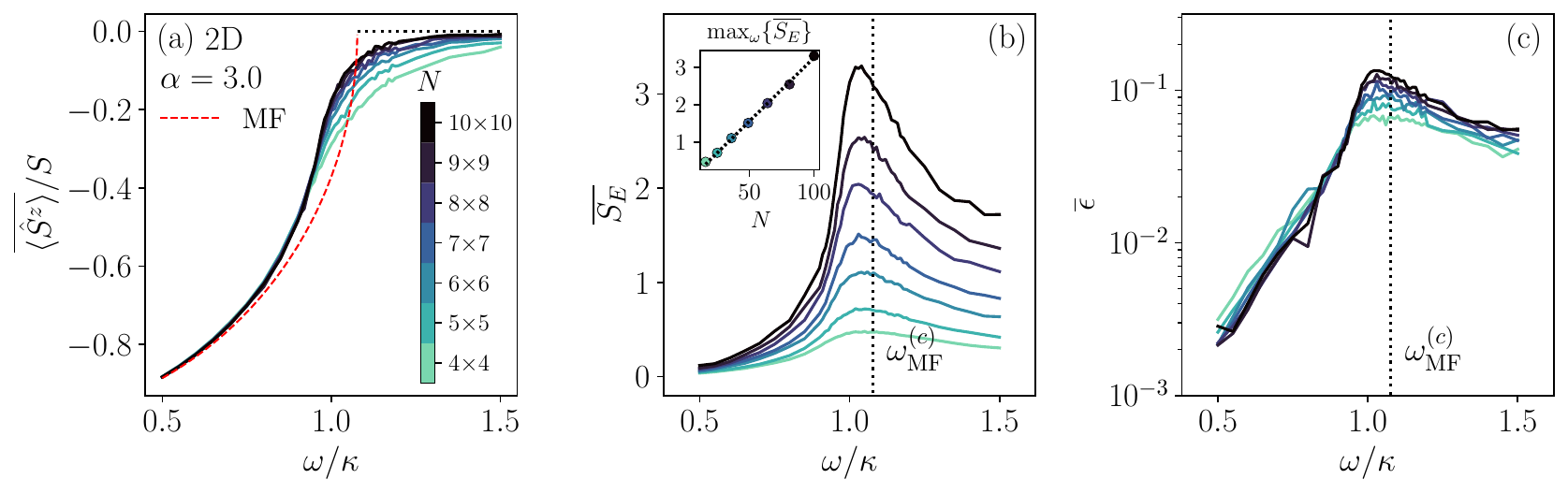}
        \caption{{Same as Fig.~\ref{fig:btc-zz-a02} but for $\alpha=3.0\,$.} \hl{The dotted line in the inset is a linear fit for $N$ ranging from 25 to 100 and we report the coefficient of determination to be $R^2=0.9983$.}}
        \label{fig:btc-zz-a3}
    \end{figure*}

    \subsection{{Results for the power-law spin model in 2D}}\label{app:results-2d}
\begin{figure}
        \centering
        \includegraphics[width=\linewidth]{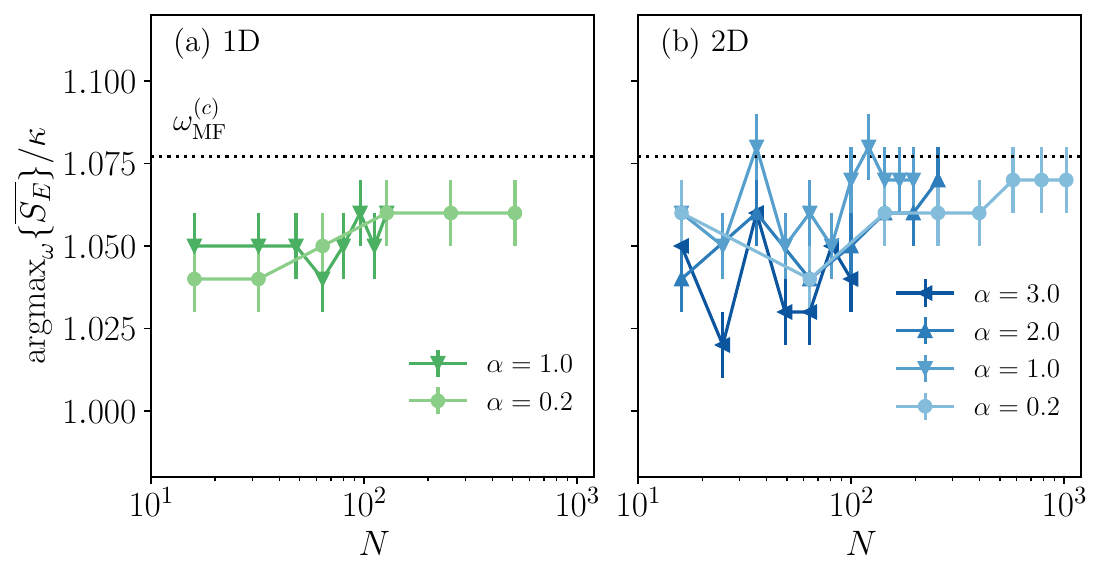}
        \caption{{The drive that maximizes the steady-state trajectory-averaged entanglement entropy, $\mathrm{argmax}_\omega\{\overline{S_E}\}$, as a function of the system size $N$ for different interaction ranges $\alpha$. The errorbar is set at the increment at which we scan $\omega$, i.e. $\Delta\omega=0.01\kappa$ in our settings. The dotted horizontal line marks the mean-field critical drive value $\omega_{\mathrm{MF}}^{(c)}$.
        }}
        \label{fig:scaling-argmax}
    \end{figure}

    {
    In this section, we present additional results on the power-law spin model for a 2D square lattice. Fig.~\ref{fig:btc-zz-a02} shows the steady-state magnetization [panel (a)], trajectory-averaged half-system entanglement entropy [panel (b)], and the spin-wave density [panel (c)], for the case of $\alpha=0.2$. As the system size $N=L^2$ increases, the magnetization shows the (dissipative) phase transition predicted by the mean-field theory. The entanglement also develops a log divergence at the same critical point, accompanied by a maximal spin-wave density (with respect to $\omega$) that decreases with $N$. As the interaction range is sufficiently long, these results are similar to the 1D case with $\alpha=0.2$ discussed in the main text (Sec.~\ref{sec:zz-long-range}), and both instances are showing essentially the mean-field (infinite-range) physics. Note that we report results for up to $N=32\times32 = 1024$ interacting spins in 2D.

    Fig.~\ref{fig:btc-zz-a2} shows the results for $\alpha=2.0$, which is the boundary case between our identification of long-range and short-range regimes. Similar to the 1D case with $\alpha=1.0$, the entanglement scaling at its maximum becomes asymptotically volume-law. The peak spin-wave density no longer vanishes with $N$, yet still remains bounded and small ($<0.1$), suggesting the validity of the spin-wave approximations. The transition in $\langle\hat{S}^z\rangle$ remains qualitatively similar to the mean-field solution (since this transition is driven by the infinite-range dissipation and not the coherent interaction term), with a small deviation originating from the quantum fluctuations close to the transition. These features are more manifest in Fig.~\ref{fig:btc-zz-a3}, where we study a moderately short-range case with $\alpha=3.0\,$. The entanglement exhibits a volume-law scaling again and the magnetization shows more deviation from the mean-field solution due to the increased quantum fluctuations. Contrary to the previous cases, the spin-wave density increases with the system size $N$ for drive values close to and beyond the mean-field critical point. We therefore limit the system sizes studied (up to $N=10\times 10 = 100$ spins in 2D) such that the spin-wave density remains small around $\overline{\epsilon}\lesssim 0.1$.

    Finally, we show in Fig.~\ref{fig:scaling-argmax} the drive value at which maximum entanglement is achieved, i.e. $\mathrm{argmax}_\omega\{\overline{S_E}\}$, for the different cases we considered in both 1D and 2D. As the system size $N$ increases, this drive value becomes asymptotically independent of $N$ and converges to a value close to the mean-field critical point.
    }

 \subsection{\hl{Benchmark of the binary approximation of the Gaussian noise}}\label{sec:bin-benchmark}
\begin{figure*}[ht]
		\centering
		\includegraphics[width=\linewidth]{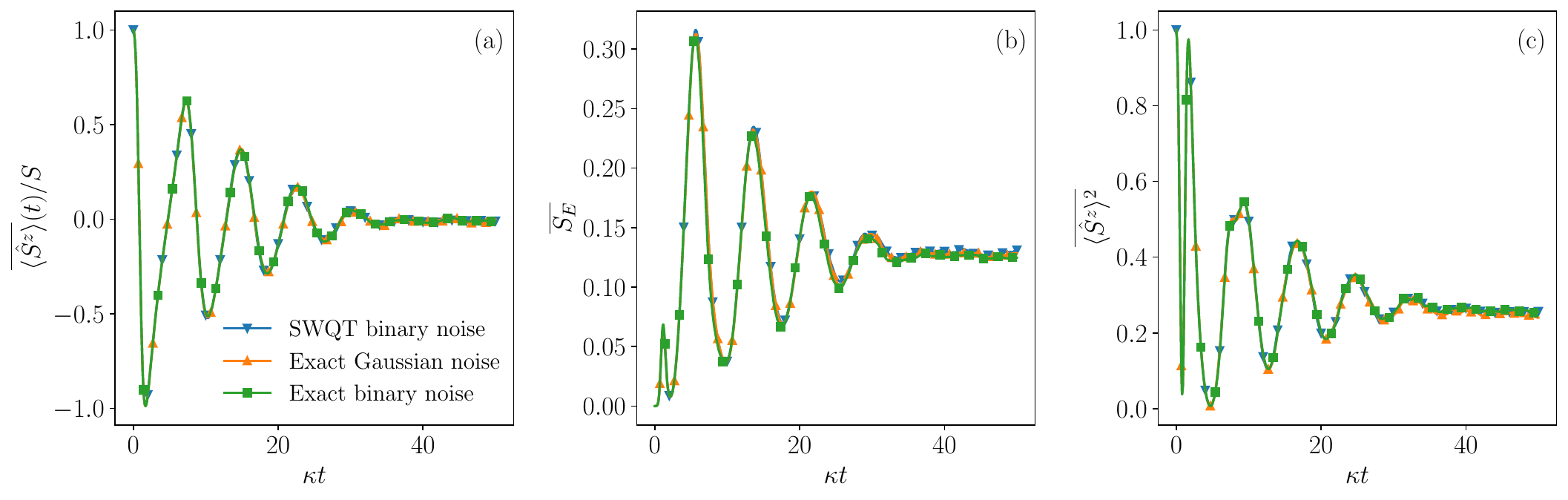}
		\caption{\hl{Benchmark of the binary approximation of the Gaussian noise in the heterodyne-detection unraveling. We compare trajectory-averaged quantities obtained from different methods and different types of noise (see legend), namely 1) the spin-wave quantum trajectory (SWQT) method with binarized noise, 2) the exact solution of the stochastic master equation with Gaussian noise and 3) the exact solution but with binary noise replacing the gaussian noise. Panel (a) compares the results on a linear quantity $\overline{\langle\hat{S}^z\rangle}$. Panels (b) and (c) show the benchmark on two different nonlinear quantities, the half-system entanglement entropy $\overline{S_E}$ and the squared magnetization $\overline{\langle\hat{S}^z\rangle^2}$ respectively. Parameters: $\omega=1.3\kappa$, $J = 0.1\kappa$, $\alpha=0$ and $S=64$.}}
		\label{fig:bivar-bench}
	\end{figure*}

\hl{
In this section, we verify the validity of the binary approximation of the Gaussian noise presented in the Methods section by computing the trajectory-averaged quantities including both linear and non-linear functions of the state, and by comparing the results with the exact solution obtained with the Gaussian noise. An example of the benchmark is shown in Fig.~\ref{fig:bivar-bench}, where we computed the magnetization $\overline{\langle\Shat^z\rangle}$, the half-system entanglement entropy $\overline{S_E}$ and the nonlinear quantity $\overline{\langle\Shat^z\rangle^2}$ from three different methods: 1) Spin-wave quantum trajectories with binarized noise, 2) exact solution of the stochastic master equation with Gaussian noise and 3) exact solution with the binary approximation of the noise. In all cases, we observe excellent agreement among the results obtained in these different ways.
}

	\section{Generalization to spin-boson systems}\label{sec:spin-boson-theory}

	\begin{figure*}[t]
		\centering
		\includegraphics[width=\linewidth]{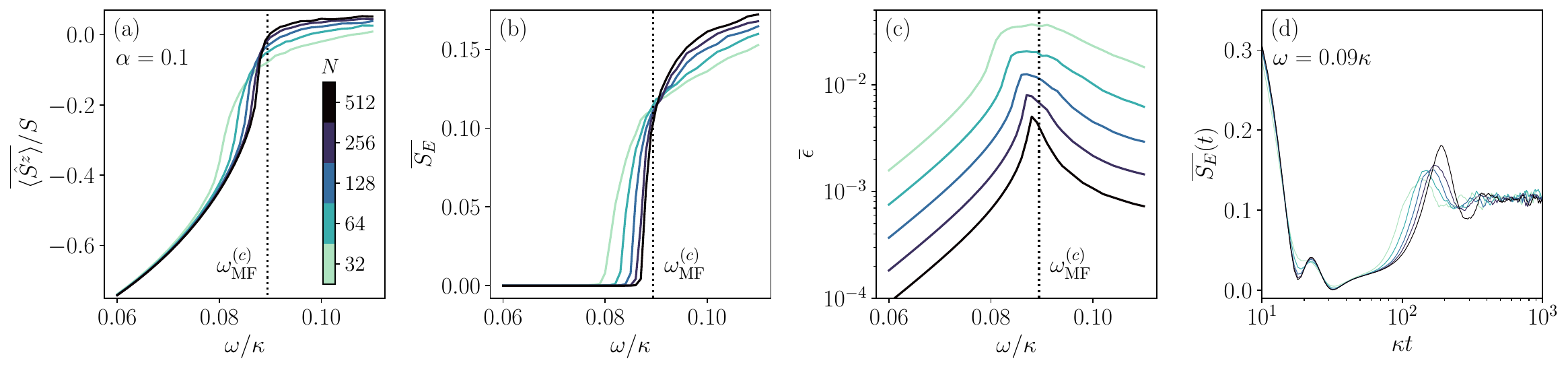}
		\caption{Results for the power-law interacting Tavis-Cummings model with $\alpha=0.1$, $J=0.01\kappa$ and $\lambda=0.2\kappa$ of different system sizes (colorbar shared across all panels). (a) Steady-state average $z$ magnetization as a function of the drive $\omega$. The vertical dotted line marks the critical point predicted by the mean-field theory $\omega^{(c)}_{\mathrm{MF}}\simeq 0.089\kappa$. (b) Steady-state of the trajectory-averaged entanglement entropy (between the spins and the cavity) as a function of $\omega$. (c) Steady-state spin-wave density. (d) Dynamics of the trajectory-averaged spin-cavity entanglement entropy for a driving value $\omega=0.09\kappa$, in linear-log scale. }
		\label{fig:spin-boson-zz-long}
	\end{figure*}

	In this Section, we demonstrate a possible extension of our spin-wave theory, which allows the investigation of spin-boson systems. We adopt the same Gaussian approximation for the bosonic mode as for the bosonized spins, such that the state of the entire system can be characterized by the first and second moments, requiring a quadratic number of variables to specify the Gaussian variational ansatz in the general case. For illustration purposes, let us consider a driven-dissipative Tavis-Cummings model with long-range spin interactions. The spins are collectively driven and resonantly coupled to a single-mode cavity (with annihilation operator $\aaa$ for the cavity bosonic mode), and the cavity undergoes single-photon dissipation. The Hamiltonian can be written as follows,
	
	\bea
	\Hhat =&~ \omega\Shat^x+\dfrac{2sJ}{\ncal}\sum_{i\neq j}\dfrac{\sigmaz_i\sigmaz_j}{{\lVert\vec{r}_i-\vec{r}_j\rVert}^\alpha}\\&+\dfrac{\lambda}{\sqrt{2S}}(\daaa \Shat^- + \aaa\Shat^{+})\,,
	\eea
	which is essentially the spin Hamiltonian~(7) in the main text with an additional term (in the second line) describing the spin-boson coupling (with strength $\lambda$). 
	The average dynamics of the system can be described by the Lindblad master equation,
	\bea
	\dfrac{\d}{\dt}\rhohat = -\rmi[ \Hhat,\rhohat ] + \kappa\dcal[\aaa](\rhohat)\,,
	\eea
	where $\kappa$ represents the loss rate of the cavity photons. 
	
	Let us consider the same quantum-state diffusion unraveling as in the main text. For a time-independent operator $\ohat$, the stochastic evolution of its expectation is given by
	\bea
	\d\langle\ohat\rangle ={}&\rmi \dt\left\langle[ \Hhat,\ohat ]\right\rangle + \kappa\dt\left\langle\daaa\ohat\aaa-\dfrac{1}{2}\left\{\daaa\aaa,\ohat\right\}\right\rangle\\
	&+ \sqrt{\kappa}\left\{ \d Z^*\left(\langle\ohat\aaa\rangle-\langle\ohat\rangle\langle\aaa\rangle \right)+ \right. \\ &\phantom{====.}\left. \d Z\left( \langle\daaa\ohat\rangle-\langle\daaa\rangle\langle\ohat\rangle \right) \right\}\,,
	\eea
	where the single-channel noise $\d Z$ satisfies $\d Z^2=0$ and $|\d Z|^2=\d t$.
	This corresponds to monitoring the cavity output field with a heterodyne-detection scheme.
	\subsection{Equations of motion}
	
	As our model preserves the translational symmetry of the state on the level of single trajectories, within the Gaussian approximation for both the cavity and the spins, we only need to keep track of the following quantities to fully specify the state:
	\begin{itemize}
		\item First moments: $\alpha\equiv\langle\aaa\rangle$, $\beta\equiv\langle\bbb_{n_0}\rangle$ (where $n_0$ is an arbitrary spin index, whose value has no importance due to the translational invariance of the spins.).
		\item Second moments:
		\begin{itemize}
			\item Photon-photon correlations: \\$\ua\equiv\langle\delhata\delhata\rangle$, $\va\equiv\langle\ddelhata\delhata\rangle$.
			\item Photon-spin correlations:\\$\uab\equiv\langle\delhata\delhatb_{n_0}\rangle$, $\vab\equiv\langle\ddelhata\delhatb_{n_0}\rangle$.
			\item Spin-spin correlations:\\$\ub_m\equiv\langle\delhatb_{n_0}\delhatb_{{n_0}+m}\rangle$, $\vb_m\equiv\langle\ddelhatb_{n_0}\delhatb_{{n_0}+m}\rangle$.
		\end{itemize}
	\end{itemize}
	Here, we define $\delhata\equiv\aaa-\langle\aaa\rangle$ and $\delhatb_i\equiv\bbb_i-\langle\bbb_i\rangle$, which are time-dependent operators. The spin correlators satisfy $\ub_m=\ub_{-m}$ and $\vb_m=\vb_{-m}$ due to spatial reflection symmetry in the considered model.
	
	In the re-alignment step, as the rotation applies only to the spin operators, the cavity mode is not affected. Therefore, after obtaining the angular increments $\Delta\theta$ and $\Delta\phi$ from the incremented $\beta$ using Eq.~(28) in the main text, we have the following update rules,
	\bea
	\beta&\leftarrow0\,,\quad\\\uab&\leftarrow\uab\rme^{-\rmi\Delta\phi\cos\theta}\,,\quad\\\vab&\leftarrow\vab\rme^{-\rmi\Delta\phi\cos\theta}\,,\quad\\\ub_m&\leftarrow\ub_m\rme^{-2\rmi\Delta\phi\cos\theta}\,,
	\eea
	and the other correlators remain unchanged.

	\subsection{Results on the driven-dissipative long-range Tavis-Cummings model}
	
	Note that in the regime where the cavity dissipation is much faster than any other time scale in the system, i.e. when $\omega,J,\lambda\ll\kappa$, the cavity mode can be adiabatically eliminated and this model reduces effectively to the long-range spin model with collective spin dissipation $\Shat^-$ as considered in the main text, since the lossy cavity acts as a Markovian reservoir for the spin system that quickly evacuates entropy into the environment via the spin-boson interaction. To illustrate the generalization of our spin-wave quantum trajectory method to spin-boson systems, we place ourselves close to this regime, such that results qualitatively reminiscent of those in the spin-only model as presented in the main text can be expected. We fix the parameters as $s=1/2$, $\lambda=0.2\kappa$, $J=0.01\kappa$ and $\alpha=0.1$, and {consider a 1D spin chain in the cavity}. The results are shown in Fig.~\ref{fig:spin-boson-zz-long}. The spin magnetization [panel (a)] displays similar behavior as compared to Fig.~\ref{fig:btc-zz-long} (a) of the main text, where a continuous dissipative phase transition predicted by the mean-field theory emerges. This transition is also accompanied by an entanglement phase transition [panel (b)] for the entanglement between the spins and the cavity, from an unentangled phase to an entangled one as the driving increases. The spin-wave density [panel (c)] displays a maximum close to the critical point, while the peak value decreases with the system size $N$, which is a signature of the long-range nature of the system. Finally, the spin-cavity entanglement also exhibits fast saturation, as shown in panel (d) for the case close to criticality. These results suggest a possible implementation of the long-range spin model with collective dissipation (as considered in the main text) in atom-cavity platforms, and that the brute-force experimental detection of the entanglement phase transition can also be realized with mitigated post-selection overhead.

\end{document}

%% file: shortcuts.tex
\renewcommand{\(}{\left(}
\renewcommand{\)}{\right)}
\renewcommand{\[}{\left[}
\renewcommand{\]}{\right]}

\let\originalleft\left
\let\originalright\right
\renewcommand{\left}{\mathopen{}\mathclose\bgroup\originalleft}
\renewcommand{\right}{\aftergroup\egroup\originalright}

\newcommand{\vect}[1]{\boldsymbol{#1}}
\renewcommand{\vec}[1]{\vect{#1}}

\makeatletter
\newcommand*\bigcdot{{\color{gray}\mathpalette\bigcdot@{1.}}}
\newcommand*\bigcdot@[2]{\mathbin{\vcenter{\hbox{\scalebox{#2}{$\m@th#1\bullet$}}}}}
\makeatother

\def\rme{{\rm {e}}}
\def\rmi{{\rm {i}}}
\renewcommand{\d}{{\rm {d}}}
\def\dt{{\rm {d}}t}

\def\tr{{\rm{Tr}}}

\newcommand{\alphatil}{{\Tilde{\alpha}}}

\newcommand{\delhat}{\hat{\delta}}
\newcommand{\ddelhat}{\hat{\delta}^\dagger}

\newcommand{\rhohat}{\hat{\rho}}

\newcommand{\sigmam}{\hat{\sigma}^{-}}
\newcommand{\sigmap}{\hat{\sigma}^{+}}

\newcommand{\sigmaz}{\hat{\sigma}^{z}}

\newcommand{\sigmahat}{\hat{\sigma}}

\newcommand{\schr}{Schr\"{o}dinger}

\newcommand{\aaa}{\hat{a}}
\newcommand{\daaa}{\hat{a}^\dagger}

\newcommand{\Ahat}{\hat{A}}

\newcommand{\bbb}{\hat{b}}
\newcommand{\dbbb}{\hat{b}^\dagger}

\newcommand{\BS}{\widehat{\mathrm{BS}}}

\newcommand{\dcal}{\mathcal{D}}

\newcommand{\Ehat}{\hat{E}}

\newcommand{\Hhat}{\hat{H}}

\newcommand{\Khat}{\hat{K}}

\newcommand{\Lhat}{\hat{L}}

\newcommand{\Lcal}{\mathcal{L}}

\newcommand{\mvec}{{\vec{m}}}

\newcommand{\ncal}{\mathcal{N}}

\newcommand{\ohat}{\hat{O}}

\newcommand{\ocal}{\mathcal{O}}

\newcommand{\Ohat}{\hat{O}}

\newcommand{\phat}{\hat{p}}

\newcommand{\rvec}{\vec{r}}

\newcommand{\rf}{{\mathrm{RF}}}

\newcommand{\Shat}{\hat{S}}
\newcommand{\shat}{\hat{s}}

\newcommand{\Uhat}{\uhat}
\newcommand{\uhat}{\hat{U}}

\newcommand{\ua}{u^{(a)}}

\newcommand{\uab}{u^{(ab)}}

\newcommand{\ub}{u^{(b)}}

\renewcommand{\va}{v^{(a)}}

\newcommand{\vab}{v^{(ab)}}

\renewcommand{\vb}{v^{(b)}}

\newcommand{\delhata}{\hat{\delta}^{(a)}}
\newcommand{\delhatb}{\hat{\delta}^{(b)}}

\newcommand{\ddelhata}{\hat{\delta}^{(a)\dagger}}
\newcommand{\ddelhatb}{\hat{\delta}^{(b)\dagger}}

\newcommand{\xhat}{\hat{x}}

\newcommand{\xtil}{{\Tilde{x}}}

\newcommand{\ytil}{{\Tilde{y}}}

\newcommand{\ztil}{{\Tilde{z}}}

\newcommand{\pmx}[1]{\begin{pmatrix}
		#1
\end{pmatrix}}

\newcommand{\bea}{\begin{equation}\begin{aligned}}
		\newcommand{\eea}{\end{aligned}\end{equation}}
\newcommand{\be}{\begin{equation}}
	\newcommand{\ee}{\end{equation}}